\documentclass[superscriptaddress, aps, prx, reprint, nofootinbib]{revtex4-2}

\usepackage{graphicx}% Include figure files
\usepackage{amsmath,amssymb,mathtools,booktabs,multirow}
\usepackage{dcolumn}% Align table columns on decimal point
\usepackage{bm}% bold math
\usepackage{xcolor}
\usepackage{soul}
\usepackage{braket}
\usepackage[separate-uncertainty=true,multi-part-units=single]{siunitx}
\usepackage{scalerel}
\usepackage{enumitem}
\usepackage{bbm}

\definecolor{Mathematica1}{rgb}{0.368417, 0.506779, 0.709798}
\definecolor{Mathematica2}{rgb}{0.880722, 0.611041, 0.142051}
\definecolor{Mathematica3}{rgb}{0.560181, 0.691569, 0.194885}
\definecolor{darkred}{rgb}{0.545,0,0}
\definecolor{dullblue}{rgb}{0,0.298,0.49}
\definecolor{blue3}{RGB}{31, 119, 180}
\usepackage[colorlinks=true
,urlcolor=dullblue
,anchorcolor=blue3
,citecolor=dullblue
,filecolor=blue3
,linkcolor=darkred
,menucolor=blue3
,pagecolor=blue3
,linktocpage=true
,pdfproducer=medialab
]{hyperref}

\newcommand{\dd}{\mathop{\mathrm{d}\!}{}}

\renewcommand{\vec}[1]{\boldsymbol{\mathbf{#1}}}

\DeclareMathOperator\erf{erf}

\newcommand{\ped}[1]{_{\mathrm{#1}}}

\newcommand{\sped}[1]{_{\textsc{#1}}}

\DeclarePairedDelimiter{\abs}{\lvert}{\rvert}

\def\beq{\begin{equation}}
\def\eeq{\end{equation}}

\begin{document}

\title{Self-acceleration of Hardening Binaries}
% Force line breaks with \\

\author{Giovanni Maria Tomaselli}
\email{tomaselli@ias.edu}
\affiliation{School of Natural Sciences, Institute for Advanced Study, Princeton, NJ 08540, USA}

\author{Thomas F.~M.~Spieksma}
\email{thomas.spieksma@physics.ox.ac.uk}
\affiliation{Rudolf Peierls Centre for Theoretical Physics, University of Oxford, Parks Road, Oxford, OX1 3PU, United Kingdom}

\begin{abstract}
A Keplerian binary immersed in a bath of lighter particles hardens by ejecting them through gravitational slingshots. This process drives, for example, the evolution of supermassive black hole binaries following galaxy mergers, and has long been described with just two parameters: the hardening rate and the eccentricity growth rate. Here we show that the secular dynamics is substantially richer. Combining symmetry arguments with extensive three-body scattering experiments, we demonstrate that the medium exerts a net force on the binary's center of mass (CoM), induces apsidal precession, and rotates the orbital plane when the CoM velocity has an out-of-plane component. Remarkably, these deterministic effects persist even in a perfectly uniform and isotropic medium, as the binary’s own asymmetry provides the propulsion. The interplay of self-acceleration, precession, and dynamical friction drives the CoM along an outward spiral. For supermassive black hole binaries, this displacement dominates over Brownian motion and approaches the radius of influence, suggesting they may be significantly offset from their host galaxies' centers. The displacement also enlarges the stellar loss cone, with direct implications for the final-parsec problem. We further show that the previously reported circularization of small-mass-ratio binaries is a numerical artifact of truncating long-lived encounters: all binaries undergo eccentricity growth. Our results enrich the standard picture of binary hardening and have implications in a variety of astrophysical contexts, including gravitational-wave source populations.
\end{abstract}

\makeatletter
\g@addto@macro\@FMN@list{\@TBN@opr{0}{\hspace*{-1em}This work was developed jointly by the authors.\vspace{2.5pt}}}
\makeatother

\maketitle

%%%%%%%%%%%%%%%%%%%%%%%
\section{Introduction}
%%%%%%%%%%%%%%%%%%%%%%%
\emph{How does a Keplerian binary evolve due to its interaction with a background of lighter particles?} This question arises in a broad range of astrophysical contexts, such as supermassive black hole (SMBH) binaries in a stellar background, intermediate mass-ratio binaries in dense star clusters, and stellar-mass binaries embedded in accretion disks.
Among these examples, the evolution of SMBH binaries following galaxy mergers has been studied most extensively. There, the binary's journey from kiloparsec scales to coalescence is commonly described in three phases \cite{Begelman:1980vb,Merritt:2004gc}. First, dynamical friction against the background brings the SMBHs closer together, until they form a hard binary. Below a much smaller separation, gravitational-wave (GW) emission becomes efficient and drives the final inspiral and merger. In between, the binary is too tight for dynamical friction yet too wide for GWs to merge it in a Hubble time.
Shrinking in this intermediate regime is thought to be dominated by \emph{gravitational slingshots} \cite{Saslaw:1974}: stars passing close to the binary are ejected at high velocity, carrying away energy and angular momentum. Beginning with the seminal work by Quinlan~\cite{Quinlan:1996vp}, this slingshot process has been studied through three-body scattering experiments. Quinlan introduced two key evolution parameters: the \emph{hardening rate} $H$, which governs the shrinkage of the semi-major axis $a$, and the \emph{eccentricity growth rate} $K$, which tracks the evolution of the eccentricity $e$. Together, $H$ and $K$ determine the coupled evolution of $a$ and $e$, and have been the main output of numerous subsequent studies \cite{Sesana:2006xw,Sesana:2006ne,Sesana:2007vr,Sesana:2011wf,Sesana:2015haa,Rasskazov_2019,Bonetti:2020iwk,Gualandris:2022kxh} that have refined the numerical methods and extended the parameter space.
In this work, we revisit the problem and show that it is considerably richer than previously appreciated. From symmetry arguments and scattering experiments, we find that, in general, three-body slingshots exert a \emph{net average force} on the binary's center of mass (CoM). This can be seen as a form of \emph{self-acceleration} of the binary: even if the medium is uniform and isotropic, the asymmetry of the binary itself generates the propulsion, unless it is exactly circular or of equal mass. This is a \emph{deterministic}, \emph{secular} effect: the CoM is pushed in a specific direction within the orbital plane, determined by the eccentricity vector, mass ratio, and angular momentum direction. Furthermore, we show that slingshots induce a \emph{precession} of the longitude of periapsis. Finally, if the CoM has a nonzero out-of-plane velocity component, the orbital plane undergoes a \emph{rotation}. Following the definitions of $H$ and $K$, we embed the acceleration, precession and rotation in three new dimensionless parameters, $\vec P$, $Q$, and $\vec R$, respectively. These three effects had not been previously appreciated in the literature on binary hardening. The asymmetry in the ejection pattern of scattered stars was noticed in~\cite{Sesana:2006xw,Rasskazov_2019}, but the recoil on the binary itself was not considered.
These effects should be clearly distinguished from the Brownian motion of massive binaries, which arises from the discrete, stochastic nature of stellar encounters. Brownian motion comprises both the \emph{translational} random walk of the CoM and the \emph{rotational} evolution of the orbital plane and eccentricity vector \cite{Merritt:2000yg,Merritt:2001hy,Chatterjee:2002sq,Bortolas:2016lge,Khan:2020,Varisco_2021}. The effects we identify are qualitatively different, as they persist in the smooth limit where Brownian motion vanishes. Previous studies of SMBH ``binary wandering'' have focused only on these stochastic effects and their possible role in refilling the loss cone \cite{Merritt:2000yg,Merritt:2001hy,Chatterjee:2001ty,Chatterjee:2002zz,Milosavljevic:2002bn,Chatterjee:2002sq,Chatterjee:2003ji,Laun:2004uq,Berczik:2005ff,Bortolas:2016lge,DiCintio:2020}.
The hardening of a binary immersed in an environment is sometimes approximated by applying Chandrasekhar's dynamical friction formula \cite{Chandrasekhar:1943ys}. This approach does not return physically correct results, and the error has been known for a long time \cite{1983AJ.....88.1269H,Quinlan:1996vp}. Nevertheless, using this method, the authors of Ref.~\cite{Cardoso:2020lxx} noticed for the first time (to our knowledge) that an asymmetric hardening binary experiences a nonzero CoM force. Although we rely on entirely different methods, Ref.~\cite{Cardoso:2020lxx} can thus be considered as a precursor of our work, as far as the self-acceleration is concerned.
Chandrasekhar's dynamical friction does, however, play a role in our problem. Once the binary's CoM self-accelerates, the deflection of distant background particles forces it to settle into a terminal velocity. This is a friction on the binary \emph{as a whole}, rather than on its individual components. We show that the interplay between the CoM force, dynamical friction and the slingshot-induced precession leads to a circling or outward-spiraling motion of the CoM. For SMBH binaries, the radius of this trajectory is a significant fraction of the binary's radius of influence---much larger than the Brownian random walk previously considered.
Our results rely on a large suite of three-body scattering experiments, comprising 20 million simulations for every choice of mass ratio and eccentricity, spanning a wide range of binary parameters for a total of several hundred combinations. Unlike previous studies, we store the results in a way that allows all evolution parameters to be computed for arbitrary nonzero CoM velocity, through a reweighting of the background velocity distribution.
In addition, we carry out the first systematic study of the effect of the integration-time cutoff on the evolution parameters $H$, $K$, $\vec P$, $Q$, and $\vec R$. Notably, when the cutoff is chosen large enough, we find that all binaries---regardless of mass ratio and eccentricity---undergo eccentrification. We explicitly demonstrate that the circularization of small-mass-ratio binaries previously reported in the literature \cite{Rasskazov_2019,Bonetti:2020iwk} is a numerical artifact arising from the premature truncation of long-lived encounters.
Our findings are relevant for numerous astrophysical applications. One of the most immediate questions, for example, is the extent to which this newly identified CoM motion contributes to refilling the stellar loss cone for SMBH binaries, thus alleviating the ``final parsec problem'' \cite{Milosavljevic:2001vi,Milosavljevic:2002bn}. While we attempt a qualitative discussion of this and other implications, we expect proper in-depth analyses to require dedicated studies. Due to the general character of the problem, however, we use neutral language throughout the paper, phrasing it in terms of a binary hardening in a background of \emph{particles}---we only choose an astrophysical setting and physical timescales in a dedicated section.

\begin{figure}
\centering
\includegraphics[width=0.48\textwidth]{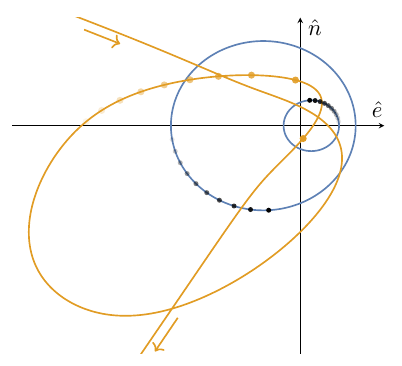}
\caption{Visualization of the slingshot of a particle off a binary with $q=0.3$ and $e=0.4$. The particle comes in from the top-left, enters a short-lived bound orbit, and, after a close encounter with the more massive binary component, is ejected at high velocity toward the bottom left. The variation of the particle's momentum induces a recoil on the binary.}
\label{fig:scattering}
\end{figure}

The remainder of the paper is organized as follows. Section~\ref{sec:hardening} sets up the problem, defines the evolution parameters and uses symmetry arguments to argue which of them are (or are not) expected to vanish. Section~\ref{sec:three-body} describes our scattering experiments and the numerical extraction of the evolution parameters, whose results are presented in Sec.~\ref{sec:results}. Section~\ref{sec:Tmax} discusses the effect of the integration-time cutoff and the physics of slingshots off small-mass-ratio binaries. We then solve for the binary's evolution in Sec.~\ref{sec:binary_evolution}, and discuss astrophysical applications in Sec.~\ref{sec:astro_impl}. We conclude in Sec.~\ref{sec:conclusions}. Several appendices contain additional details: Appendix~\ref{app:convergence} presents numerical convergence tests; Appendix~\ref{app:bmin} discusses the effective minimum impact parameter for dynamical friction; Appendix~\ref{app:comparison} studies eccentricity growth at small mass ratios and compares our results with those of Refs.~\cite{Rasskazov_2019,Bonetti:2020iwk}; Appendix~\ref{app:uncertainties} explains how we determine the uncertainties quoted in Sec.~\ref{sec:binary_evolution}; and Appendix~\ref{app:Nbody} validates our results with a direct $N$-body approach.
Our code, as well as the data obtained from our three-body scattering experiments, is publicly available on GitHub \cite{BBH_stars:github}.
%%%%%%%%%%%%%%%%%%%%%%%%%%%
\section{Binary hardening}
\label{sec:hardening}
%%%%%%%%%%%%%%%%%%%%%%%%%%%
We now set up the general framework for studying the interaction of a Keplerian binary with a background of lighter particles. We aim to keep the formalism as general as possible, deferring specific astrophysical applications to Sec.~\ref{sec:astro_impl}.

%%%%%%%%%%%%%%%%%%%%%%%%%%%%%%%%%%%%%%%%
\subsection{Binary in a uniform medium}
\label{sec:binary-uniform-medium}
%%%%%%%%%%%%%%%%%%%%%%%%%%%%%%%%%%%%%%%%
Consider a Keplerian binary of total mass $M$ immersed in a background of point-like particles, each of mass $m\ll M$. Far from the binary, the background is assumed to be uniform and isotropic, with mass density $\rho = n m$, where $n$ is the number density. As a natural choice for a virialized or thermalized background, we take the velocity distribution to be Maxwellian with one-dimensional velocity dispersion $\sigma$. The probability density function (p.d.f.) for each Cartesian velocity component is
\beq
f\sped{1d}(v_x)=\frac{e^{-v_x^2/(2\sigma^2)}}{\sqrt{2\pi}\sigma}\,,
\label{eqn:maxwellian-1d}
\eeq
with analogous expressions for $v_y$ and $v_z$. The p.d.f.\ for the velocity magnitude $v$ is
\beq
f(v)=\sqrt{\frac2\pi}\frac{v^2}{\sigma^3}e^{-v^2/(2\sigma^2)}\,.
\label{eqn:maxwellian}
\eeq
We work in the limit $m\to 0$ and $n\to\infty$ at fixed $\rho$, which makes the background smooth and the problem non-stochastic, allowing us to focus on the \emph{average} effect of each interaction between a particle and the binary. Stochastic aspects of similar problems have been studied elsewhere \cite{Merritt:2000yg,Merritt:2001hy,hamilton2024eccentricity}.
The binary components have masses
\beq
M_1=\frac{M}{1+q}\,,\qquad M_2=\frac{qM}{1+q}\,,
\eeq
where $q\le1$ is the mass ratio, and the reduced mass is $\mu=qM/(1+q)^2$. The orbit is characterized by the semi-major axis $a$ and eccentricity $e$. Its orientation is fixed by two vectors: the angular momentum $\vec L=L\hat L$, and the eccentricity vector $\vec e=e\hat e$, which points from $M_1$ toward $M_2$ when they are at periapsis. Furthermore, we denote by $\vec V$ the velocity of the binary's CoM with respect to the background.
The problem is thus fully specified by the following parameters:
\begin{itemize}
\item background parameters: \{$\rho$, $\sigma$\};
\item binary parameters: \{$M$, $q$, $a$, $e$, $\hat L$, $\hat e$, $\vec V$\}.
\end{itemize}
Finally, it is useful to define the hard binary radius,
\beq
a_h\equiv\frac{G\mu}{4\sigma^2}\,,
\label{eqn:a_h}
\eeq
which separates two qualitatively different regimes. For $a\gtrsim a_h$, close binary-particle interactions are typically quick impulsive encounters. For $a\lesssim a_h$, they are instead chaotic three-body interactions.
%%%%%%%%%%%%%%%%%%%%%%%%%%%%%%%%%%%%%
\subsection{Symmetry considerations}
\label{sec:symmetry}
%%%%%%%%%%%%%%%%%%%%%%%%%%%%%%%%%%%%%
We now use symmetry arguments to determine which aspects of the binary's evolution are, in principle, permitted.  We assume that the component masses, and hence $M$ and $q$, do not change over time. The parameters that can evolve are therefore $a$, $e$, $\hat L$, $\hat e$, and $\vec V$, and their rates of change must be expressible in terms of these same parameters, together with the constants $\rho$, $\sigma$, $M$, and $q$.
As we shall see, this analysis reveals that, in addition to the well-known hardening and eccentricity growth, the binary generically experiences other effects that have been entirely overlooked in the literature: a net force on its center of mass, an apsidal precession, and a rotation of its orbital plane.
%%%%%%%%%%%%%%%%%%%%%%%%%%%%%%%
\subsubsection{Binary at rest}
\label{sec:symmetry-rest}
%%%%%%%%%%%%%%%%%%%%%%%%%%%%%%%
We begin with the case $\mathbf{V}=0$, where the binary is at rest with respect to the background.
\indent \emph{Semi-major axis.} The semi-major axis $a$ is a scalar, and its evolution is not protected by any symmetry. Given the power $\dot E$ transferred by the particles to the binary,
\beq
\dot E=-\frac{GM\mu}2\frac{\dd}{\dd t}\bigg(\frac1a\bigg)\,,
\eeq
we define, following Quinlan~\cite{Quinlan:1996vp}, the dimensionless \emph{hardening rate},
\beq
H\equiv\frac{\sigma}{G\rho}\frac{\dd}{\dd t}\bigg(\frac1a\bigg)=-\frac{2\sigma\dot E}{G^2M\mu\rho}\,.
\label{eqn:H}
\eeq
\indent \emph{Angular momentum and eccentricity.} The angular momentum $\vec L$, with magnitude $L=\sqrt{GM\mu^2a(1-e^2)}$, is the only pseudovector within the set of parameters. The torque $\vec\tau$ on the binary is also a pseudovector, therefore it must be parallel to $\hat L$,
\beq
\vec\tau\equiv \frac{\dd\vec L}{\dd t}=\tau_{\hat L}\hat L\,.
\eeq
As an immediate consequence of this, we see that the orbital plane cannot change: $\dd\hat L/\dd t=0$. Furthermore, the out-of-plane component of the torque, $\tau_{\hat L}\equiv\dd L/\dd t$, governs the evolution of the eccentricity. Once again, consistent with previous literature, we express this in terms of a dimensionless \emph{eccentricity growth rate},
\beq
K\equiv-a\frac{\dd e}{\dd a}=-\frac{1-e^2}{2e}+\frac{\sqrt{1-e^2}}{2e}\sqrt{\frac{GM}{a^3}}\frac{\tau_{\hat L}}{\dot E}\,.
\label{eqn:K}
\eeq
The parameters $H$ and $K$ were first introduced by Quinlan \cite{Quinlan:1996vp}, and have since been studied by numerous authors \cite{Sesana:2006xw,Sesana:2006ne,Sesana:2007vr,Sesana:2011wf,Gualandris:2011ck,Sesana:2015haa,Rasskazov:2016gzt,Rasskazov_2019,Bonetti:2020iwk,Berczik:2020awp,Varisco_2021,Gualandris:2022kxh}. However, they do not completely determine the evolution of the binary, nor its interaction with the environment. We now show that this is the case by continuing with our symmetry arguments.
\indent \emph{Eccentricity vector and apsidal precession.} When $\vec V=0$, the set of available parameters contains two true vectors, namely $\hat e$ and $\hat n \equiv \hat L\times\hat e$, which together span the binary's orbital plane. Since $\hat e$ is a unit vector, it can only evolve by rotating about $\hat L$,
\beq
\frac{\dd\hat e}{\dd t}=\dot\varpi\,\hat n\,,
\eeq
leading to a precession of the binary's longitude of periapsis $\varpi$. This secular, deterministic precession is driven by the average effect of three-body scatterings. It should not be confused with the stochastic rotational Brownian motion of the eccentricity vector studied in, e.g.,~\cite{Merritt:2001hy}. 
\indent \emph{Center-of-mass force.} The velocity $\vec V$ is a true vector, with unconstrained norm. Given that the true vectors available are $\hat e$ and $\hat n$, the medium can exert a net force on the binary, directed within the orbital plane:
\beq
\vec F=M\frac{\dd\vec V}{\dd t}=F_{\hat e}\hat e+F_{\hat n}\hat n\,,
\label{eqn:F}
\eeq
By contrast, the out-of-plane component vanishes: $F_{\hat L}=0$.

This is a simple yet remarkable result, and the central finding of this work: a Keplerian binary at rest in a perfectly uniform, isotropic medium \emph{self-accelerates}. The direction of this force is determined entirely by the binary's internal structure---its eccentricity vector and mass ratio---without the need for any density gradient or anisotropy of the environment.
Special cases with enhanced symmetry impose additional constraints. A circular binary ($e=0$) enjoys full rotational symmetry about $\hat L$, which forces $\vec F=0$. Likewise, an equal-mass binary ($q=1$) is symmetric under rotations of $\SI{180}{\degree}$ about $\hat L$, which also requires $\vec F=0$. For all other configurations, that is, whenever $q<1$ and $e>0$, the CoM force is generically nonzero.
\emph{Dimensionless evolution parameters.} Similar to $H$ and $K$, we introduce dimensionless quantities that capture the variation of the CoM velocity $\vec V$ and longitude of periapsis $\varpi$ per logarithmic change of the semi-major axis. We define the \emph{acceleration parameter}
\beq
\vec P\equiv-\frac{a}\sigma\frac{\dd\vec V}{\dd a}=-\frac{G\mu}{2\sigma a}\frac{\vec F}{\dot E}=\frac{\vec F}{GM\rho aH}\,,
\label{eqn:P}
\eeq
and the \emph{precession parameter}\footnote{We define the instantaneous precession rate $\dot\varpi$ in the natural frame of the binary, where by definition its inclination vanishes. If the binary's orbital plane changes with time, as will be the case in Sec.~\ref{sec:symmetry-arbitrary}, the evolution of the orbital angles with respect to a fixed frame can be tracked by rotating into the binary frame, using $Q$ to determine their rate of change, and rotating back to the fixed frame.}
\beq
Q\equiv-a\frac{\dd\varpi}{\dd a}=-\frac{GM\mu}{2a}\frac{\dot\varpi}{\dot E}=\frac{\sigma\dot\varpi}{G\rho aH}\,.
\label{eqn:Q}
\eeq
The five parameters $H$, $K$, $P_{\hat e}$, $P_{\hat n}$, and $Q$ fully specify the instantaneous evolution of a binary at rest with respect to the background. Of these, $H$ and $K$ have been extensively studied in the literature, while $P_{\hat e}$, $P_{\hat n}$, and $Q$ are identified here for the first time.
%%%%%%%%%%%%%%%%%%%%%%%%%%%%%%%%%%%%%%%%%%%%%%%%%%%%%%%
\subsubsection{Binary moving within its orbital plane}
\label{sec:motion-orbital-plane}
%%%%%%%%%%%%%%%%%%%%%%%%%%%%%%%%%%%%%%%%%%%%%%%%%%%%%%%
We now relax the assumption of vanishing CoM velocity and consider the case where $\vec V$ lies in the orbital plane. The additional parameter $\vec V$ introduces two new pseudovectors, $\hat V\times\hat e$ and $\hat V\times\hat n$. However, both of them point along $\hat L$, and therefore the argument constraining $\vec\tau$ to be parallel to $\hat L$ remains unchanged. Similarly, no new true vector pointing in the $\hat L$ direction is available, so the force $\vec F$ is confined to the orbital plane. 
A binary initially moving within its orbital plane will thus be forever constrained to move within it: no force component along $\hat L$ can be generated, and the torque can only change the magnitude of $\vec L$ but not its direction. The dimensionless parameters governing the evolution are the same as in the case at rest: $H$, $K$, $P_{\hat e}$, $P_{\hat n}$, and $Q$.
%%%%%%%%%%%%%%%%%%%%%%%%%%%%%%%%%%%%%%%%
\subsubsection{Arbitrary binary motion}
\label{sec:symmetry-arbitrary}
%%%%%%%%%%%%%%%%%%%%%%%%%%%%%%%%%%%%%%%%
When $\vec V$ has a nonzero component $V_{\hat L}$ along $\hat L$ (i.e., perpendicular to the orbital plane), the symmetry constraints are relaxed and new effects become possible. The pseudovectors $\hat V\times\hat e$ and $\hat V\times\hat n$ now have components in the orbital plane, allowing the torque to acquire nonvanishing components $\tau_{\hat e}$ and $\tau_{\hat n}$. As a result, $\hat L$ is free to rotate,
\beq
\frac{\dd\hat L}{\dd t}=\frac{\tau_{\hat e}}L\hat e+\frac{\tau_{\hat n}}L\hat n\equiv\vec\Omega\times\hat L\,,
\eeq
where $\vec\Omega=(-\tau_{\hat n}\hat e+\tau_{\hat e}\hat n)/L$ is the instantaneous angular velocity of the orbital plane's rotation. For arbitrary binary motion, we thus introduce the \emph{rotation parameter}
\beq
\vec R\equiv-a\,\hat L\times\frac{\dd\hat L}{\dd a}=\frac{\sigma\vec\Omega}{G\rho aH}\,,
\label{eqn:R}
\eeq
which is always directed within the orbital plane.
Furthermore, because $\hat e$, $\hat n$, and $\hat V$ now span the full space, the CoM force $\vec F$ can point in any direction, and so $P_{\hat L}$ is not necessarily zero. The full set of evolution parameters when $V_{\hat L}\ne0$ is then composed of eight quantities, namely $H$, $K$, $P_{\hat e}$, $P_{\hat n}$, $P_{\hat L}$, $Q$, $R_{\hat e}$, and $R_{\hat n}$. Although this general case is more involved, we will show in Sec.~\ref{sec:evol_nonzero_vel} that the out-of-plane velocity component is generally damped over time, which allows the evolution to be well described by the simpler in-plane case of Sec.~\ref{sec:motion-orbital-plane}.
Despite the reduced symmetry, we can still draw useful conclusions about the parity properties of the evolution parameters as a function of $V_{\hat L}$. Under a parity transformation (i.e., an inversion of spatial coordinates), scalars and pseudovectors remain invariant, while true vectors pick up a minus sign. Hence, the $\hat e$ and $\hat n$ components of true vectors, as well as the $\hat L$ component of pseudovectors, are unchanged; conversely, the $\hat e$ and $\hat n$ components of pseudovectors, as well as the $\hat L$ component of true vectors, flip sign. In particular, $(V_{\hat e},V_{\hat n},V_{\hat L})\to(V_{\hat e},V_{\hat n},-V_{\hat L})$, while the other binary parameters stay unchanged. We thus find that $H$, $K$, $P_{\hat e}$, $P_{\hat n}$, and $Q$ are even functions of $V_{\hat L}$, while $R_{\hat e}$, $R_{\hat n}$, and $P_{\hat L}$ are odd functions of $V_{\hat L}$. We provide a summary of all the evolution parameters, together with their definitions and symmetry properties, in Tab.~\ref{tab:parameters}.

\begin{table*}
\centering
\begin{tabular*}{0.85\textwidth}{@{\extracolsep{\fill}}cccccc}
\toprule
\textbf{Parameter name} & \textbf{Symbol} & \textbf{Definition} & \textbf{Equation} & \textbf{Parity with} $V_{\hat L}$ & \textbf{New or known?}\\
\midrule
hardening & $H$ & $(\sigma/G\rho)\dd(1/a)/\dd t$ & \eqref{eqn:H} & even & known \cite{Quinlan:1996vp,Sesana:2006xw}\\
\midrule
eccentricity growth & $K$ & $-a\dd e/\dd a$ & \eqref{eqn:K} & even & known \cite{Quinlan:1996vp,Sesana:2006xw}\\
\midrule
\multirow{3}{*}{acceleration} & $P_{\hat e}$ & \multirow{3}{*}{$-(a/\sigma)\dd\vec V/\dd a$} & \multirow{3}{*}{\eqref{eqn:P}} & even & \multirow{3}{*}{new}\\
& $P_{\hat n}$ & & & even & \\
& $P_{\hat L}$ & & & odd & \\
\midrule
precession & $Q$ & $-a\dd\varpi/\dd a$ & \eqref{eqn:Q} & even & new\\
\midrule
\multirow{2}{*}{rotation} & $R_{\hat e}$ & \multirow{2}{*}{$-a\,\hat L\times\dd\hat L/\dd a$} & \multirow{2}{*}{\eqref{eqn:R}} & odd & \multirow{2}{*}{new}\\
& $R_{\hat n}$ & & & odd & \\
\bottomrule
\end{tabular*}
\caption{Summary of the evolution parameters. The hardening and eccentricity growth parameters $H$ and $K$ have been identified and studied in a number of previous works, starting with \cite{Quinlan:1996vp,Sesana:2006xw}. The rotation, acceleration and precession parameters $\vec P$, $Q$, and $\vec R$ are introduced here for the first time. When $V_{\hat L}=0$ (i.e., the binary's CoM has no out-of-plane velocity component), all parameters odd in $V_{\hat L}$ vanish, while all parameters even in $V_{\hat L}$ are generically nonzero.}
\label{tab:parameters}
\end{table*}

%%%%%%%%%%%%%%%%%%%%%%%%%%%%%%%%%%%%%%%%%%%%%%
\subsection{CoM force and dynamical friction}
\label{sec:df}
%%%%%%%%%%%%%%%%%%%%%%%%%%%%%%%%%%%%%%%%%%%%%%
The presence of a nonzero CoM force $\vec F$, introduced in Sec.~\ref{sec:symmetry-rest}, might at first glance seem surprising when the binary does not move ($\vec V=0$). Conversely, the case $\vec V\ne0$ is perhaps more intuitive: like any other moving object, the binary experiences \emph{dynamical friction} (acting on its CoM) due to its interaction with the environment \cite{Chandrasekhar:1943ys}. Correctly calculating and interpreting the acceleration parameter $\vec P$ requires understanding the relation between the two forces.
Dynamical friction is generally discussed for a moving point particle, so let us temporarily collapse the binary to a single point with mass $M$ and velocity $\vec V$. If the environment consisted of particles at rest ($\sigma=0$), then in the binary's frame their scattering along hyperbolic orbits would induce a drag force \cite{Binney_Tremaine2008}
\beq
\vec F_{\textsc{df},\vec V}=-\frac{4\pi(GM)^2\rho\,\vec V}{V^3}\log\Lambda(V)\,,
\label{eqn:Fdf-V}
\eeq
where 
\beq
\log\Lambda(V)\equiv\frac12\log\bigg(\frac{b\ped{max}^2+b_{90}(V)^2}{b\ped{min}^2+b_{90}(V)^2}\bigg)\,.
\label{eqn:logLambda}
\eeq
Here, $b_{90}(V)\equiv GM/V^2$ is the impact parameter for which an incoming particle is deflected by $\SI{90}{\degree}$, and $b\ped{max}$ ($b\ped{min}$) are the maximum (minimum) impact parameters of the incoming particles. Equation~\eqref{eqn:Fdf-V} is finite as $b\ped{min}\to0$, but has an infrared divergence as $b\ped{max}\to\infty$, due to the infinite extent of the medium. In reality, this is cured by the finiteness of real astrophysical systems, which is usually taken into account by replacing $b\ped{max}$ with, e.g., the typical size of the environment. Because at large distances a binary is seen by the background particles as a single object, the infrared divergence in our problem remains identical to the familiar case of a point particle.
Averaging over the Maxwellian velocity distribution in the rest frame of the medium gives
\beq
\vec F\sped{df}=\int\dd^3v\,f(\abs{\vec V-\vec v})\,\vec F_{\textsc{df},\vec v}\,.
\label{eqn:Fdf-int}
\eeq
If $\log \Lambda$ is taken to be a constant, it can be pulled out of the integral, and Eq.~\eqref{eqn:Fdf-int} reduces to the well-known Chandrasekhar's formula for dynamical friction,
\beq
\vec F\sped{df}=-\frac{4\pi(GM)^2\rho\,\vec V}{V^3}\bigg(\erf\chi-\frac{2\chi}{\sqrt\pi}e^{-\chi^2}\bigg)\log\Lambda\,,
\label{eqn:chandrasekhar}
\eeq
where $\chi=V/(\sqrt2\sigma)$.
For our purposes, however, it is useful to work with the exact expression \eqref{eqn:Fdf-int}, where $\log\Lambda$ depends on $v$. It will also be convenient to let $b\ped{min}$ depend on the velocity of the incoming particle, $b\ped{min}\to b\ped{min}(v)$, in a way that we explain in detail in Sec.~\ref{sec:three-body}. We then define the \emph{dynamical friction parameter},
\beq
\vec P\sped{df}\equiv\frac{\vec F\sped{df}}{GM\rho aH}\,,
\label{eqn:Pdf}
\eeq
which quantifies the CoM force induced by the particles whose impact parameter falls between $b\ped{min}(v)$ and $b\ped{max}$.
Because \eqref{eqn:Pdf} treats the binary as a point mass, it is only accurate if $b\ped{min}(v)$ is large enough that the background particles remain at distances much larger than $a$ from the binary's CoM. We thus calculate the total force as follows: for $b>b\ped{min}(v)$, we use \eqref{eqn:Pdf} to compute $\vec P\sped{df}$, while for $b<b\ped{min}(v)$ we perform three-body scattering experiments (see Sec.~\ref{sec:three-body}), which correctly resolve the binary, and determine their contribution $\vec P\sped{3bs}$. The total acceleration parameter is then the sum of the two contributions,
\beq
\vec P=\vec P\sped{3bs}+\vec P\sped{df}\,.
\eeq
We therefore \emph{glue} together the effects of the scatterings studied numerically with those of the scatterings that we can approximate analytically. There is no sharp conceptual separation between the two terms: the split is dictated by computational cost and is implemented through an appropriate choice of $b\ped{min}(v)$.
The CoM force $\vec F$ discussed in this paper can thus be seen as a generalization of $\vec F\sped{df}$ from the case of a point particle to the case of a Keplerian binary. When $\vec V=0$, the dynamical friction contribution $\vec F\sped{df}$ vanishes, but the three-body scattering contribution $\vec F\sped{3bs}$ does not.
%%%%%%%%%%%%%%%%%%%%%%%%%%%%%%%%%
\section{Three-body experiments}
\label{sec:three-body}
%%%%%%%%%%%%%%%%%%%%%%%%%%%%%%%%%
In the previous section, we used symmetry arguments to establish how the binary can, in principle, evolve, and defined the dimensionless parameters $H$, $K$, $\vec P$, $Q$, and $\vec R$ that govern such evolution. We now describe how we compute these parameters numerically by performing a large number of three-body scattering experiments (code available at \cite{BBH_stars:github}).
We study the restricted three-body problem: a massless test particle is scattered off a binary whose components follow a fixed Keplerian orbit, according to the geometry defined in Sec.~\ref{sec:binary-uniform-medium} and illustrated in Fig.~\ref{fig:scattering}. For each scattering event, we record the changes in the particle's energy, angular momentum, and velocity, as well as the induced precession of the binary's longitude of periapsis. These changes are averaged over many scatterings at fixed incoming speed $v$, and then weighted over the Maxwellian velocity distribution~\eqref{eqn:maxwellian} to yield the evolution parameters as functions of $a/a_h$.
%%%%%%%%%%%%%%%%%%%%%%%%%%%%%%%%%%%
\subsection{Numerical integration}
\label{sec:num-int}
%%%%%%%%%%%%%%%%%%%%%%%%%%%%%%%%%%%
Consider a particle approaching the binary with asymptotic speed $v$ (i.e., its speed infinitely far away from the binary). We describe below our numerical setup for the integration of the particle's trajectory. Most of these choices follow standard practice in three-body scattering experiments (e.g.,~\cite{Quinlan:1996vp,Sesana:2006xw,Rasskazov_2019,Bonetti:2020iwk}).
\begin{itemize}
\item \textbf{Initial position.} We consider a sphere of radius $r\ped{sph}=50a$ centered on the binary's CoM. The particle's initial position is randomly sampled uniformly on that sphere.
\item \textbf{Initial speed.} The particle's initial speed is set to $v\ped{sph}=\sqrt{v^2+2GM/r\ped{sph}}$, such that its asymptotic speed far from the binary would be $v$ if the binary were collapsed to a point particle with mass $M$.
\item \textbf{Impact parameter.} We only consider particles whose initial orbit would bring them within a pericenter distance $r_p \leq 5a$ from the origin, if the binary were treated as a point mass. This corresponds to a \emph{maximum} impact parameter
\beq
b\ped{cut}^2(v)=r_p^2\bigg(1+\frac{2GM}{v^2r_p}\bigg)\,.
\label{eqn:bmin-v}
\eeq
We sample the particle's impact parameter $b$ such that $b^2$ is uniformly distributed in $[0,b\ped{cut}^2]$, which corresponds to a uniform flux of incoming particles.

Particles passing farther away contribute to dynamical friction, and we treat them analytically as described in Sec.~\ref{sec:df}. To \emph{glue} the two contributions together, we thus identify $b\ped{cut}$ with the \emph{minimum} impact parameter of particles contributing to dynamical friction, which amounts to imposing $b\ped{min}(v)=b\ped{cut}(v)$ in Eq.~\eqref{eqn:Pdf}.
\item \textbf{Velocity direction.} The azimuthal angle of the particle's initial velocity with respect to the radial direction is sampled uniformly in $[0, 2\pi]$.
\item \textbf{Binary's phase.} The initial time of integration is sampled uniformly in $[0, T]$, where
\beq
T = 2\pi\sqrt{\frac{a^3}{GM}}
\label{eq:binary_period}
\eeq
is the binary's orbital period.  This randomizes the binary's phase at the start of each scattering.
\end{itemize}
We integrate the equations of motion of the restricted three-body problem using an explicit Runge--Kutta (\texttt{RK45}) method with an adaptive time step. The step-size control is designed to properly resolve all portions of the particle's orbit, and in particular the close encounters with either binary component. To further regularize strong close encounters, we soften the gravitational attraction as
\beq
\frac{\vec r}{r^3} \to \frac{\vec r}{(r^2+\epsilon^2)^{3/2}}\,,
\label{eq:softening}
\eeq
with $\epsilon=10^{-5}a$. We have verified that this small softening length is sufficient to remove numerical artifacts from the results. Additional numerical checks including convergence tests and comparisons with different integrators are discussed in Appendix~\ref{app:convergence}.
We stop the integration of each scattering event when either of the following two conditions is satisfied:
\begin{itemize}
\item the particle reaches a distance $r>r\ped{sph}$ from the origin with speed greater than $\sqrt{2GM/r\ped{sph}}$, that is, the particle is outgoing and unbound (if the binary is treated as a point particle);
\item the integration time reaches a maximum cutoff $T\ped{max}$.
\end{itemize}
The first case corresponds to ``resolved'' scatterings, which contribute to the measurement of  the evolution parameters $H$, $K$, $\vec P$, $Q$, $\vec R$. In the second case, the particle has become temporarily bound to the binary in an extremely long-lived orbit. We discard such events for the purpose of the computation.
Unless otherwise stated, we adopt $T\ped{max}=10^{11}T/(2\pi)$. We will usually write this in terms of the parameter $N\ped{max}\equiv T\ped{max}/T=10^{11}/(2\pi)$. This very large cutoff ensures that the vast majority of scatterings are resolved: for $v \gtrsim 10^{-2}\sqrt{GM/a}$ (roughly corresponding to $a/a_h \gtrsim 10^{-2}$), typically only 1 or 2 particles in $10^4$ are discarded. This number, however, rapidly grows for very hard and very asymmetric binaries, as shown in Tab.~\ref{tab:resolved-fractions}. As we will show, this has important consequences for the evolution parameters, particularly $K$ at small mass ratios. Care must thus be taken in the choice of $T\ped{max}$. We will systematically study its impact on the evolution parameters in Sec.~\ref{sec:Tmax}.
%%%%%%%%%%%%%%%%%%%%%%%%%%%%%%%%%%%%%%%%%%%%%%%%%
\subsection{Extracting the evolution parameters}
\label{sec:extracting-evolution-parameters}
%%%%%%%%%%%%%%%%%%%%%%%%%%%%%%%%%%%%%%%%%%%%%%%%%
%%%%%%%%%%%%%%%%%%%%%%%%%%%%%%%%%%%%%%%%%%%%%%%%%%%%
\subsubsection{Binary at rest (\,$\mathrm{\bf V}=0$)}
\label{sec:weigh-maxwellian}
%%%%%%%%%%%%%%%%%%%%%%%%%%%%%%%%%%%%%%%%%%%%%%%%%%%%
For each resolved scattering, we record the changes in the particle's energy $\Delta E$, angular momentum $\Delta\vec\ell$, and asymptotic velocity $\Delta\vec v$, all measured with respect to its initial condition. In computing the asymptotic values, we approximate the binary as a point mass of mass $M$ and assume that the particle follows a Keplerian hyperbolic orbit for $r>r\ped{sph}$. In addition, we track the precession of the binary's longitude of periapsis $\varpi$. During the simulation, the binary remains on its Keplerian non-precessing orbit; however, we can compute the instantaneous perturbation to $\varpi$ induced by the presence of the particle using Gauss' planetary equation, which for vanishing inclination reads
\beq
\dot\varpi=-\sqrt{\frac{a(1-e^2)}{GMe^2}}\bigg(\cos\phi\,a_r-\frac{2+e\cos\phi}{1+e\cos\phi}\sin\phi\,a_\phi\bigg)\,,
\label{eqn:gauss}
\eeq
where $a_r$ and $a_\phi$ are the radial and tangential components of the relative acceleration that the test particle imparts on the binary, and $\phi$ is the true anomaly. The cumulative change $\Delta\varpi$ is obtained by integrating the right-hand side of \eqref{eqn:gauss} over the whole simulation.
For each value of $v$, we perform $10^4$ independent scattering experiments and compute the sample averages $\braket{\Delta E}$, $\braket{\Delta\vec\ell}$, $\braket{\Delta\vec v}$, and $\braket{\Delta\varpi}$. From these, we define the average power $\dot E_v$, torque $\vec\tau_v$, force $\vec F_v$, and precession rate $\dot\varpi_v$ induced by particles with asymptotic speed $v$ as
\begin{align}
\dot E_v  &= -\pi b\ped{cut}^2\,\rho\,v\,\frac{\langle \Delta E\rangle}{m}\,,\\[4pt]
\vec\tau_v & = -\pi b\ped{cut}^2\,\rho\,v\,\frac{\langle\Delta\vec\ell\rangle}{m}\,,\\[4pt]
\vec F_v &= -\pi b\ped{cut}^2\,\rho\,v\,\langle\Delta\vec v\rangle\,,\\[4pt]
\dot\varpi_v &=  \pi b\ped{cut}^2\,\rho\,v\,\frac{\langle\Delta\varpi\rangle}{m}\,.
\end{align}
We then repeat this procedure for many values of $v$, in order to sample the full Maxwellian velocity distribution. More specifically, given some minimum and maximum limits on the binary hardness $[a/a_h]\ped{min}$ and $[a/a_h]\ped{max}$, we select 2000 logarithmically spaced values of $v$ within the interval
\beq
\frac1{20}\frac{\sqrt{q}}{1+q}\sqrt{\Big[\frac{a}{a_h}\Big]\ped{min}}<\sqrt{\frac{a}{GM}}\,v<\frac{8\sqrt{q}}{1+q}\sqrt{\Big[\frac{a}{a_h}\Big]\ped{max}}\,,
\label{eqn:limits-v}
\eeq
chosen to properly resolve both the low-$v$ and high-$v$ tails of the Maxwellian. We thus end up with a total of $\num{2e7}$ simulations for every choice of $q$ and $e$. Throughout, we fix $[a/a_h]\ped{min}=10^{-3}$ and $[a/a_h]\ped{max}=10^2$.
To obtain the evolution parameters as a function of $a/a_h$, we weigh the results with the Maxwellian velocity distribution, given in \eqref{eqn:maxwellian-1d} and \eqref{eqn:maxwellian},
\beq
X=\int_0^\infty X_vf(v)\dd v\,,
\label{eqn:weigh-maxwellian}
\eeq
for $X\in\{\dot E,\vec\tau,\vec F,\dot\varpi\}$. These results are then folded into the hardening rate $H$, eccentricity growth rate $K$, acceleration parameter $\vec P$, precession parameter $Q$, and rotation parameter $\vec R$, according to equations \eqref{eqn:H}, \eqref{eqn:K}, \eqref{eqn:P}, \eqref{eqn:Q}, and \eqref{eqn:R}.
We determine the statistical uncertainty on $X$ by calculating the standard error of the mean associated with $\braket{\Delta Z}$, for $\Delta Z\in\{\Delta E,\Delta \vec\ell,m\Delta\vec v,-\Delta\varpi\}$, and then propagating the uncertainty with a trapezoidal rule for the integral \eqref{eqn:weigh-maxwellian}, exploiting the fact that numerical data for different $v$ are statistically independent from each other.
%%%%%%%%%%%%%%%%%%%%%%%%%%%%%%%%%%%%%%%%%%%%%%%%%%%%%
\subsubsection{Moving binary (\,$\mathrm{\bf V}\ne0$)}
\label{sec:shifted-maxwellian}
%%%%%%%%%%%%%%%%%%%%%%%%%%%%%%%%%%%%%%%%%%%%%%%%%%%%%
The procedure described above is not sufficient to handle the case where the binary's CoM moves with velocity $\vec V\ne0$ with respect to the background. The reason is that the fixed-$v$ averages $X_v$ discard information about the \emph{direction} of the incoming particle, which matters once the isotropy of the background is broken by a nonzero $\vec V$ in the binary's rest frame. We must thus keep track of the particle's initial velocity $\vec v$. 
In principle, one could retain the full information by storing the individual variations $\Delta Z\ped{\vec v}$ for every scattered particle (with $\Delta Z\in\{\Delta E,\Delta \vec\ell,m\,\Delta\vec v,-\Delta\varpi\}$) and then performing a weighted average with a shifted Maxwellian,
\beq
X=-\iiint_{-\infty}^\infty\frac{\pi b\ped{cut}^2\rho}mv\,\Delta Z_{\vec v}\,f\sped{3d}(\vec v+\vec V)\dd^3\vec v\,,
\label{eqn:X3D}
\eeq
for $X\in\{\dot E,\vec\tau,\vec F,\dot\varpi\}$, where the three-dimensional Maxwellian is
\beq
\begin{split}
f\sped{3d}(\vec v+\vec V)&=f\sped{1d}(v_x+V_x)f\sped{1d}(v_y+V_y)f\sped{1d}(v_z+V_z)\\
&=\frac{e^{-(v^2+V^2)/(2\sigma^2)}}{(2\pi\sigma^2)^{3/2}}e^{-\vec v\cdot\vec V/\sigma^2}\,.
\end{split}
\label{eqn:f3d}
\eeq
However, this brute-force approach would require storing the outcome of all $\num{2e7}$ individual scatterings for each pair $(q,e)$, which is prohibitively expensive in terms of memory.
We overcome this difficulty by decomposing $\Delta Z_{\vec v}$ in spherical harmonics,
\beq
\Delta Z_{\vec v}=\sum_{\ell m}Z^{\ell m}_vY_{\ell m}(\hat v)\,,
\label{eqn:Zvlm}
\eeq
and storing only the harmonic coefficients $Z^{\ell m}_v$ for each value of $v$, up to a maximum multipole $\ell\ped{max}=10$. This way, we only need to store $(\ell\ped{max}+1)^2$ times more information compared to the $\vec V=0$ case, as opposed to a $10^4$-fold increase from saving the data of every individual scattering. 
The distribution function \eqref{eqn:f3d} can be written with a plane wave expansion,
\beq
e^{-\vec v\cdot\vec V/\sigma^2}=\sum_{\ell m}4\pi\, i_\ell(vV/\sigma^2)Y_{\ell m}(-\hat V)Y_{\ell m}^*(\hat v)\,,
\label{eqn:plane-wave}
\eeq
where $i_\ell$ is the modified spherical Bessel function of the first kind. Plugging \eqref{eqn:Zvlm} and \eqref{eqn:plane-wave} into \eqref{eqn:X3D}, we get
\beq
\begin{split}
X=-\int_0^\infty&\frac{\pi b\ped{cut}^2\rho}m\,4\pi v^3\,\frac{e^{-(v^2+V^2)/(2\sigma^2)}}{(2\pi\sigma^2)^{3/2}}\\
&\times \sum_{\ell m}i_\ell(vV/\sigma^2)Z_v^{\ell m}Y_{\ell m}(-\hat V)\dd v.
\end{split}
\eeq
Similarly, we determine the statistical uncertainty on the integrand by expanding $\braket{(\Delta Z_{\vec v}\,f\sped{3d}(\vec v+\vec V))^2}$ in spherical harmonics as before. We then use error propagation assuming a trapezoidal rule for the $v$ integral to combine those into an uncertainty on $X$. We assign no statistical uncertainty to the dynamical friction force $\vec F\sped{df}$, so the uncertainty on $\vec P\sped{df}$ is purely due to that on $H$.

%%%%%%%%%%%%%%%%%%%%%%%%%%%%%%%%%%%%%%%%%%%%%%%
\section{Results for the evolution parameters}
\label{sec:results}
%%%%%%%%%%%%%%%%%%%%%%%%%%%%%%%%%%%%%%%%%%%%%%%
We now present our numerical results for the evolution parameters, obtained from the suite of three-body scattering simulations described in the previous section. We first consider a binary at rest with respect to the environment ($\mathbf{V} = 0$), and then examine how the parameters change when the binary's CoM moves ($\mathbf{V} \neq 0$).
%%%%%%%%%%%%%%%%%%%%%%%%%%%%%%%%%%%%%%%%%%%%%%%%%%%%%%%%%
\subsection{Evolution parameters for $\mathrm{\bf V}=0$}
%%%%%%%%%%%%%%%%%%%%%%%%%%%%%%%%%%%%%%%%%%%%%%%%%%%%%%%%%
\begin{figure*}
\centering
\includegraphics[width=\textwidth]{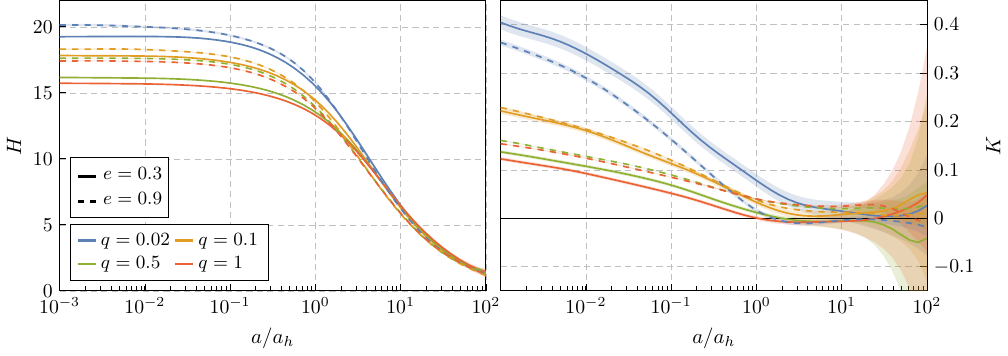}
\caption{Hardening rate $H$ and eccentricity growth rate $K$ as a function of $a/a_h$ for a number of selected values of $q$ and $e$, and vanishing CoM velocity ($\vec V=0$). The semi-transparent band around each line indicates the $1\sigma$ uncertainty. These parameters have been previously computed in the literature, e.g., in \cite{Quinlan:1996vp,Sesana:2006xw,Bonetti:2020iwk}. The agreement with these classic results provides a validation of our code.}
\label{fig:H-K}
\end{figure*}
We extract the parameters $H$, $K$, $P_{\hat e}$, $P_{\hat n}$, and $Q$ numerically following the procedure outlined in Sec.~\ref{sec:weigh-maxwellian}. The hardening rate $H$ and eccentricity growth rate $K$ have been studied previously in the literature \cite{Quinlan:1996vp,Sesana:2006xw,Rasskazov_2019,Bonetti:2020iwk}, so our results for these serve as a validation of our code. We show them in Fig.~\ref{fig:H-K} for selected values of $q$ and $e$, finding excellent agreement with, e.g., Fig.~3 and 4 of \cite{Sesana:2006xw} and Fig.~1 of \cite{Bonetti:2020iwk}. It is worth noting that our data extends down to $a/a_h=10^{-3}$, while previous literature had stopped at $a/a_h=10^{-2}$. We also study much smaller values of $q$ than most works, but we defer a detailed discussion of this aspect, and a comparison with \cite{Bonetti:2020iwk}, to Sec.~\ref{sec:Tmax} and Appendix~\ref{app:comparison}.
\begin{figure*}
\centering
\includegraphics[width=\textwidth]{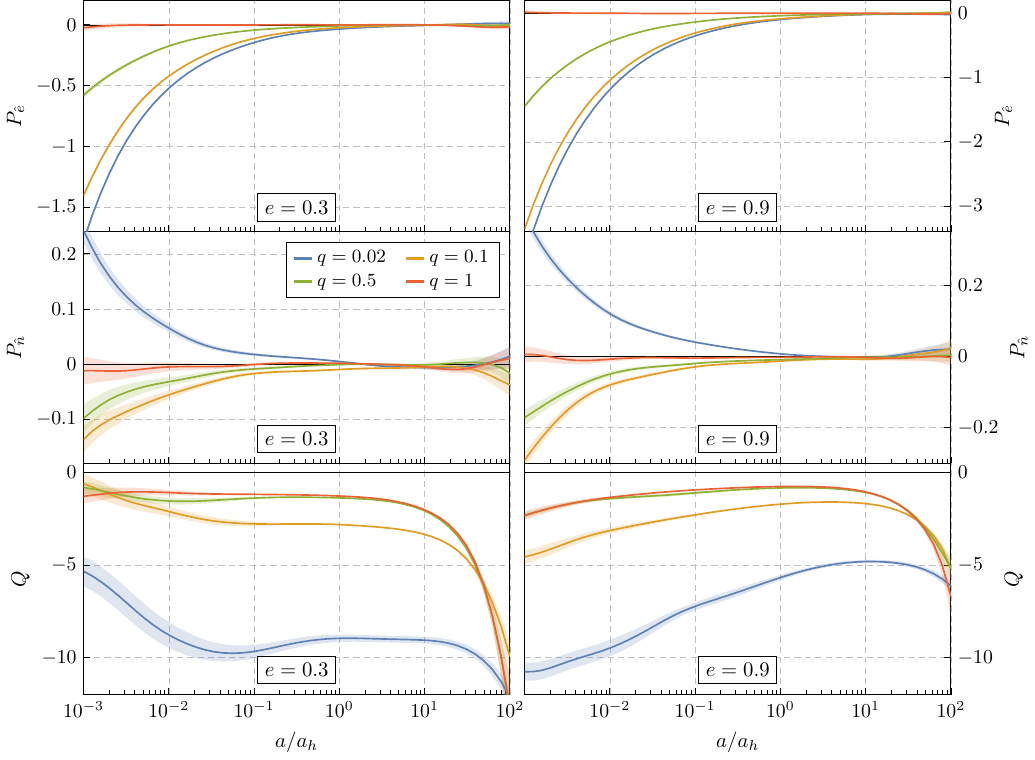}
\caption{In-plane components of the acceleration parameter $\vec P$ and the precession parameter $Q$ as a function of $a/a_h$ for selected values of $q$ and $e$, and vanishing CoM velocity ($\vec V=0$). The acceleration and precession parameters $\vec P$ and $Q$ are introduced in this work for the first time.}
\label{fig:P_xy-Q}
\end{figure*}
We plot the novel parameters, $P_{\hat e}$, $P_{\hat n}$, and $Q$, in Fig.~\ref{fig:P_xy-Q}. As expected from the symmetry arguments in Sec.~\ref{sec:symmetry}, none of these parameters is consistent with zero, except for $\vec P$ when $q=1$ or $e=0$ (we do not show results for the circular case explicitly). All parameters are generically of $\mathcal O(1)$. This means that, during one $e$-fold of binary hardening, the CoM is expected to accelerate to velocities of order $\sigma$,\footnote{As we will see in Sec.~\ref{sec:binary_evolution}, the presence of dynamical friction for $\vec V\ne0$ significantly reduces the CoM velocity in the subsequent evolution.} and the periapsis is expected to precess by an angle of $\mathcal O(1)$.
The acceleration parameter $\vec P$ grows in magnitude for harder binaries ($a/a_h\ll1$), while the precession parameter $Q$ does not change significantly. For moderate values of $q$, the CoM force is directed in the $(-\hat{e},\,-\hat{n})$ quadrant of the orbital plane; at smaller $q$ (such as $q=0.02$), it shifts to the $(-\hat{e},\,+\hat{n})$ quadrant. We provide a visualization of the direction and magnitude of the acceleration parameter $\vec P$ in Fig.~\ref{fig:orbit-grid}, for a sample of values of $q$ and $e$.

\begin{figure*}
\centering
\includegraphics{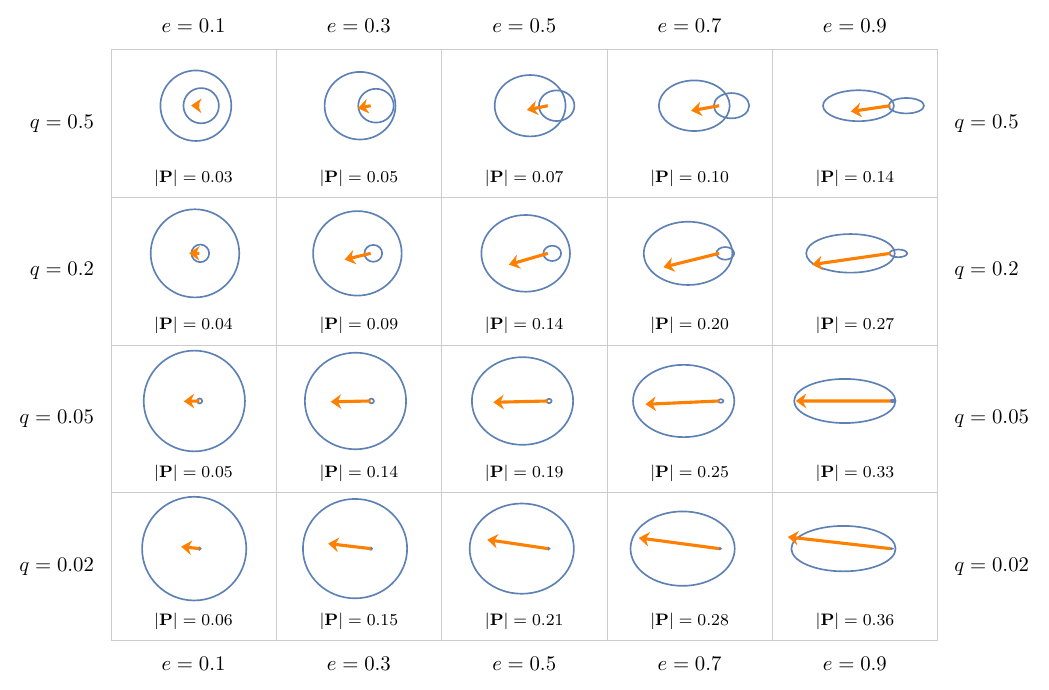}
\caption{Shape of the orbits (blue lines) and acceleration parameter $\vec P$ (orange arrows) for a sample of values of $q$ and $e$, at $\vec V=0$ and $a/a_h=0.1$. In each case, we assume that the orbital angular momentum points out of the plane.}
\label{fig:orbit-grid}
\end{figure*}

%%%%%%%%%%%%%%%%%%%%%%%%%%%%%%%%%%%%%%%%%%%%%%%%%%%%%%%%%%%
\subsection{Evolution parameters for $\mathrm{\bf V}\ne0$}
\label{sec:evol_nonzero_vel}
%%%%%%%%%%%%%%%%%%%%%%%%%%%%%%%%%%%%%%%%%%%%%%%%%%%%%%%%%%%
\begin{figure*}
\centering
\includegraphics[width=\textwidth]{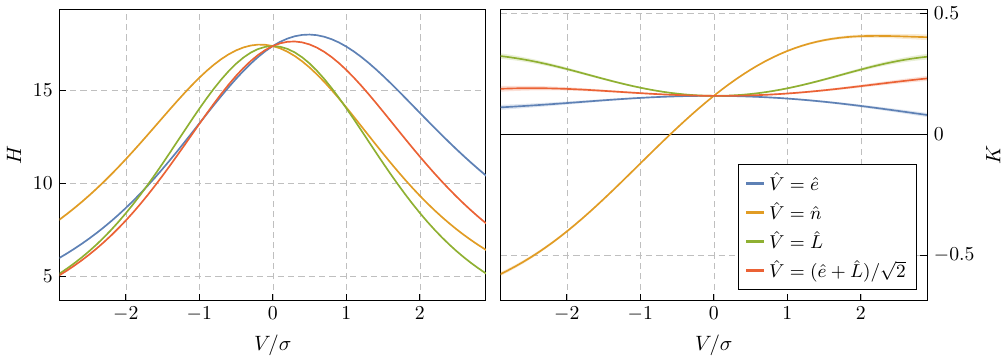}
\caption{Hardening rate $H$ and eccentricity growth rate $K$ as a function of $V=\vec V\cdot\hat V$ for $a/a_h=0.1$, $q=0.2$, $e=0.6$, and different choices for the CoM velocity direction $\hat V$.}
\label{fig:H-K-vsV}
\end{figure*}
We now examine the evolution parameters when the binary has a CoM velocity $\vec V\ne0$. Their numerical extraction follows the procedure described in Sec.~\ref{sec:shifted-maxwellian}, which involves decomposing the data into spherical-harmonic moments and weighing them with a shifted Maxwellian distribution. As argued in Sec.~\ref{sec:symmetry-arbitrary}, all eight parameters---$H$, $K$, $P_{\hat e}$, $P_{\hat n}$, $P_{\hat L}$, $Q$, $R_{\hat e}$, and $R_{\hat n}$---are generically nonzero in this case. 
Figure~\ref{fig:H-K-vsV} shows how $H$ and $K$ change when the binary's CoM moves with velocity $V$ along one of its axes, $\hat e$, $\hat n$ or $\hat L$, or along the direction $(\hat e+\hat L)/\sqrt2$. The hardening parameter $H$ reaches a maximum for $V\lesssim\sigma$, at a location that depends on the velocity direction, and then decreases with increasing $V$. For small $V$, the eccentricity growth parameter $K$ changes only to quadratic order in $V$ (in which case it remains generally bounded and positive), unless $\vec V$ has a nonzero component along $\hat n$.
\begin{figure*}
\centering
\includegraphics[width=\textwidth]{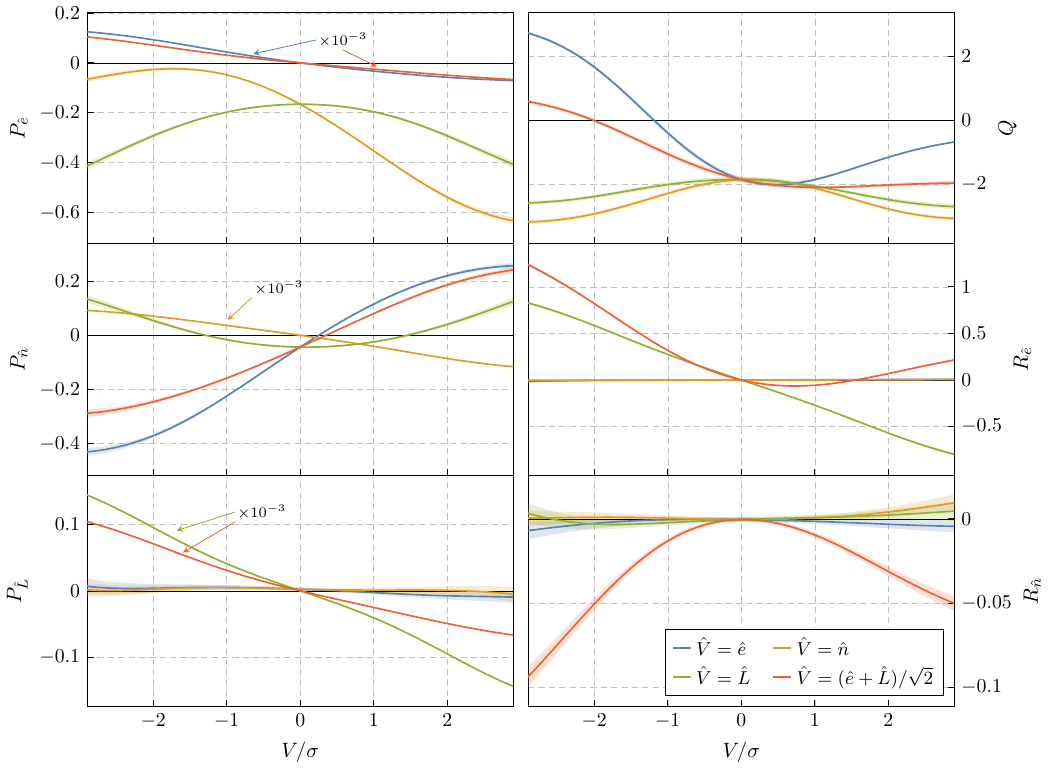}
\caption{Components of the acceleration parameter $\vec P=\vec P\sped{3bs}+\vec P\sped{df}$ (\emph{left panels}), precession parameter $Q$ and components of the rotation parameter $\vec R$ (\emph{right panels}) as a function of $V=\vec V\cdot\hat V$ for $a/a_h=0.1$, $q=0.2$, $e=0.6$, and different choices for the CoM velocity direction $\hat V$. We assume $b\ped{max}=GM/\sigma^2$ in the evaluation of $\vec P\sped{df}$. This term is generally much larger than $\vec P\sped{3bs}$, and for this reason we rescale the affected components of $\vec P$ by a factor of $10^{-3}$, as indicated in each left panel.}
\label{fig:P-RQ-vsV}
\end{figure*}
We show the $\vec V$-dependence of $\vec P$, $Q$ and $\vec R$ in Fig.~\ref{fig:P-RQ-vsV}. The parity properties of all curves with respect to $V_{\hat L}$ agree with the symmetry arguments from Sec.~\ref{sec:symmetry} and Tab.~\ref{tab:parameters}. In particular, $P_{\hat L}$, $R_{\hat e}$, $R_{\hat n}$ are compatible with zero when $\hat V = \hat e$ or $\hat V = \hat n$. For the binary parameters chosen in Fig.~\ref{fig:P-RQ-vsV}, $R_{\hat n}$ appears to vanish when $V_{\hat e}=V_{\hat n}=0$, but this is a coincidence specific to these parameters: at higher eccentricities, $R_{\hat n}$ no longer vanishes.
Of particular importance is the behavior of $\vec P$ as a function of $\vec V$. The components of $\vec P$ orthogonal to $\vec V$ are unaffected by $\vec P\sped{df}$, receiving contributions only from $\vec P\sped{3bs}$. In the example shown in Fig.~\ref{fig:P-RQ-vsV}, these components remain of $\mathcal O(1)$ for velocities $V$ of order $\sigma$. On the other hand, the components of $\vec P$ not orthogonal to $\vec V$ receive contributions from $\vec P\sped{df}$, which we evaluate in Fig.~\ref{fig:P-RQ-vsV} by fixing $b\ped{max}$ equal to the radius of influence of the binary, $b\ped{max}=r_i\equiv GM/\sigma^2$ \cite{1972ApJ...178..371P}. The dynamical friction contribution $\vec P\sped{df}$ generally dominates over $\vec P\sped{3bs}$; for this reason, we rescale the corresponding curves by a factor of $10^{-3}$ in Fig.~\ref{fig:P-RQ-vsV}, as indicated in each panel.
To understand why $\vec P\sped{df}$ dominates over $\vec P\sped{3bs}$, we evaluate Chandrasekhar's formula \eqref{eqn:chandrasekhar} for $V\sim\sigma$,
\beq
\abs{\vec F\sped{df}}\sim\frac{4\pi(GM)^2\rho}{\sigma^2}\log\Lambda\,,
\eeq
from which
\beq
\abs{\vec P\sped{df}}\sim\frac{16\pi(1+q)^2}{qH}\frac{a_h}{a}\log\Lambda\,.
\eeq
While $\vec P\sped{3bs}$ is naturally of $\mathcal O(1)$, for a hard binary ($a\ll a_h$), we must have $\abs{\vec P\sped{df}}\gg1$. Thus, although a binary at rest does accelerate due to $\vec P(0)=\vec P\sped{3bs}(0)\ne0$, we expect dynamical friction to slow down its motion well before it can reach velocities of order $\sigma$. We present a full numerical evolution of the binary parameters and its CoM motion in Sec.~\ref{sec:binary_evolution}.
%%%%%%%%%%%%%%%%%%%%%%%%%%%%%%%%%%%%%%%%%%%%%%%%%
\subsection{Dynamical friction drag coefficient}
\label{sec:drag}
%%%%%%%%%%%%%%%%%%%%%%%%%%%%%%%%%%%%%%%%%%%%%%%%%
Based on the discussion above, we expect dynamical friction to confine the binary's velocity to $V\ll\sigma$, and eliminate its out-of-plane component $V_{\hat L}$. The $\vec V=0$ results for the evolution parameters thus accurately capture the binary's evolution. The only exception is the acceleration parameter, so it is useful to investigate its small-$V$ behavior. To linear order, we can write
\beq
\vec P(\vec V)\approx\vec P(0)-\gamma\frac{\vec V}\sigma\,.
\label{eqn:P-small-V}
\eeq
Both $\vec P(0)$ and the dimensionless drag coefficient $\gamma$ are determined by our three-body scattering experiments. In principle, the coefficient $\gamma$ is a tensor, as the amount of drag depends not only on the binary's internal parameters $a/a_h$, $q$, and $e$, but also on the direction of $\vec V$ with respect to the binary's axes.
It is, however, insightful to collapse the binary to a point mass and interpret the drag term in \eqref{eqn:P-small-V} as the dynamical friction acting on it. The full expression for $\vec F\sped{df}$, given in \eqref{eqn:Fdf-int}, can be expanded to linear order in $\vec V$, yielding
\beq
\vec F\sped{df}\approx-\frac{4\sqrt{2\pi}(GM)^2\rho\,\langle\log\Lambda\rangle}{3\sigma^3}\,\vec V\,,
\label{eqn:Fdf-smallV}
\eeq
where we defined the effective Coulomb logarithm
\beq
\braket{\log\Lambda}\equiv\int_0^{\infty}e^{-v^2/2\sigma^2}\log\Lambda(v)\,\frac{v\dd v}{\sigma^2}\,.
\label{eqn:lnL-mean}
\eeq
Matching \eqref{eqn:Fdf-smallV} onto \eqref{eqn:P-small-V}, we can express the drag coefficient as
\beq
\gamma\equiv\frac{16\sqrt{2\pi}}{3H}\frac{(1+q)^2}q\frac{a_h}a\,\braket{\log\Lambda}\,.
\label{eqn:gamma}
\eeq
Extracting $\gamma$ from our simulations is thus equivalent to measuring the effective Coulomb logarithm $\braket{\log\Lambda}$. For a point particle, the integral~\eqref{eqn:lnL-mean} can be computed exactly by setting $b\ped{min}=0$. Choosing $b\ped{max}=r_i=GM/\sigma^2$, this gives $\braket{\log\Lambda}\approx0.673$.
The finite size of the binary, however, induces a slight deviation in the value of $\braket{\log\Lambda}$ from the point particle result. This can be interpreted as a nonzero \emph{effective} minimum impact parameter $b\ped{min,eff}$. On physical grounds, $b\ped{min,eff}$ should correspond to a pericenter distance of order $a$ from the binary's CoM. We already adopted a similar criterion in Sec.~\ref{sec:df} and Sec.~\ref{sec:num-int} (see Eq.~\eqref{eqn:bmin-v}) to separate particles requiring a three-body integration from those whose effect we compute analytically à la Chandrasekhar. Taking the small-$a$ limit of \eqref{eqn:bmin-v}, and assuming $b\ped{min,eff}^2\sim b\ped{cut}^2(\sigma)$, we get
\beq
b^2\ped{min,eff}\sim r_ir_p\,,
\label{eqn:bmin-sigma}
\eeq
with $r_p\sim\mathcal O(a)$ being the natural expectation. For a hard binary, this is much smaller than $b_{90}^2(\sigma)=r_i^2$, and the correction it induces in $\braket{\log\Lambda}$ is thus correspondingly small. The approximation $\braket{\log\Lambda}\approx0.673$ is thus excellent for hard binaries.
To confirm this, we use the data from our three-body scattering experiments to determine what the actual value of $b\ped{min,eff}^2$ is. We do this in Appendix~\ref{app:bmin}, where we show that $b\ped{min,eff}^2$ is indeed of order $\mathcal O(r_ia)$, in agreement with \eqref{eqn:bmin-sigma}, and that it generally takes \emph{negative} values, implying that the drag on a binary is (ever so slightly) stronger than that on a point particle.
%%%%%%%%%%%%%%%%%%%%%%%%%%%%%%%%%%%%%%%%%%%%%%%%%%%%%%
\section{Long-lived encounters and small mass ratios}
\label{sec:Tmax}
%%%%%%%%%%%%%%%%%%%%%%%%%%%%%%%%%%%%%%%%%%%%%%%%%%%%%%
In the previous section we presented the evolution parameters obtained with a large integration-time cutoff $T\ped{max}$. We now study their dependence on this cutoff, focusing on small mass ratios, where previous work reported a sign change of $K$~\cite{Rasskazov_2019,Bonetti:2020iwk}.
This issue is physically important because the sign of $K$ determines whether slingshots eccentrify or circularize the binary. We show that the previously reported circularization is a numerical artifact caused by premature truncation of long-lived encounters. When these encounters are resolved, $K$ remains positive; a detailed comparison with Refs.~\cite{Rasskazov_2019,Bonetti:2020iwk}, including a controlled reproduction of their results, is deferred to Appendix~\ref{app:comparison}.
%%%%%%%%%%%%%%%%%%%%%%%%%%%%%%%%%%%%%%%%%%%%%%
\subsection{Physics of long-lived encounters}
\label{sec:physics-longlived}
%%%%%%%%%%%%%%%%%%%%%%%%%%%%%%%%%%%%%%%%%%%%%%
At small mass ratios $q \ll 1$, the secondary acts as a weak perturber in the potential of the primary. A close interaction between the particle of mass $m$ and the binary exchanges an energy $\delta E \sim qGMm/a$. The typical kinetic energy of an incoming particle is $E\ped{kin}\sim(3/2)m\sigma^2=3qGMm/(8a_h(1+q)^2)$. The probability of a temporary capture, which occurs if $E\ped{kin}+\delta E<0$, is thus of $\mathcal{O}(1)$ for $a/a_h\ll1$, as it simply depends on the sign of $\delta E$. Furthermore, this probability is independent of $q$ for $q\ll1$, since both $E\ped{kin}$ and $\delta E$ scale linearly with $q$.
The \emph{lifetime} of a captured orbit, however, does depend on $q$. A captured particle orbits the binary on an approximately Keplerian orbit with binding energy $|E_\star| \sim \delta E \sim qGMm/a$ and semi-major axis $a_\star \sim GMm/|E_\star| \sim a/q$. Each close return imparts another kick of order $\pm\delta E$, so the particle's energy performs a one-dimensional random walk. The number of kicks needed to escape is $\mathcal{O}(1)$ and independent of $q$, but the time between successive encounters is set by
\beq
T_\star \sim T \left(\frac{a_\star}{a}\right)^{3/2} \sim T\,q^{-3/2}\,,
\eeq
where $T$ is the binary's orbital period~\eqref{eq:binary_period}. The total duration of a long-lived encounter, in units of the binary period $T$, is therefore
\beq
\label{eq:Nesc}
N_\star \equiv \frac{T_\star}{T} \sim q^{-3/2}\,.
\eeq
This scaling is consistent with the dependence of the evolution parameters on the integration-time cutoff, which we now discuss.
%%%%%%%%%%%%%%%%%%%%%%%%%%%%%%%%%%%%%%%%%%%%%%%%%%%%%%%%%%%
\subsection{Cutoff dependence of the evolution parameters}
\label{sec:Tmax-dependence}
\begin{figure*}
\centering
\includegraphics[width=\textwidth]{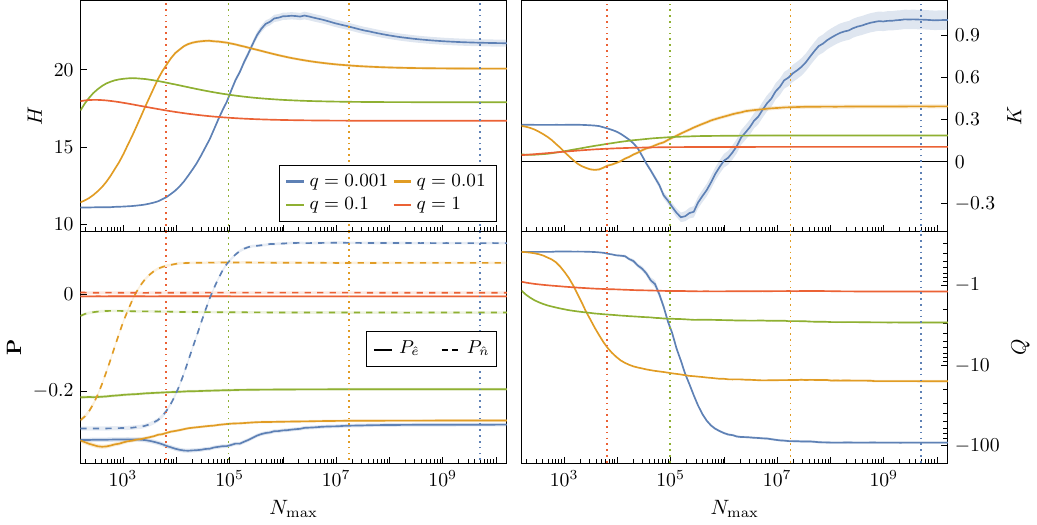}
\caption{Evolution parameters as a function of the integration-time cutoff, quantified in units of binary periods as $N\ped{max}=T\ped{max}/T$, for a hard binary with $a/a_h=0.1$, $e=0.6$, and $\vec V = 0$. Vertical dashed lines mark the hardening timescale $N\ped{hard}=T\ped{hard}/T$, given in~\eqref{eqn:Nhard}, for each mass ratio.}
\label{fig:HKPQ-Tmax}
\end{figure*}

\begin{figure*}
\centering
\includegraphics[width=\textwidth]{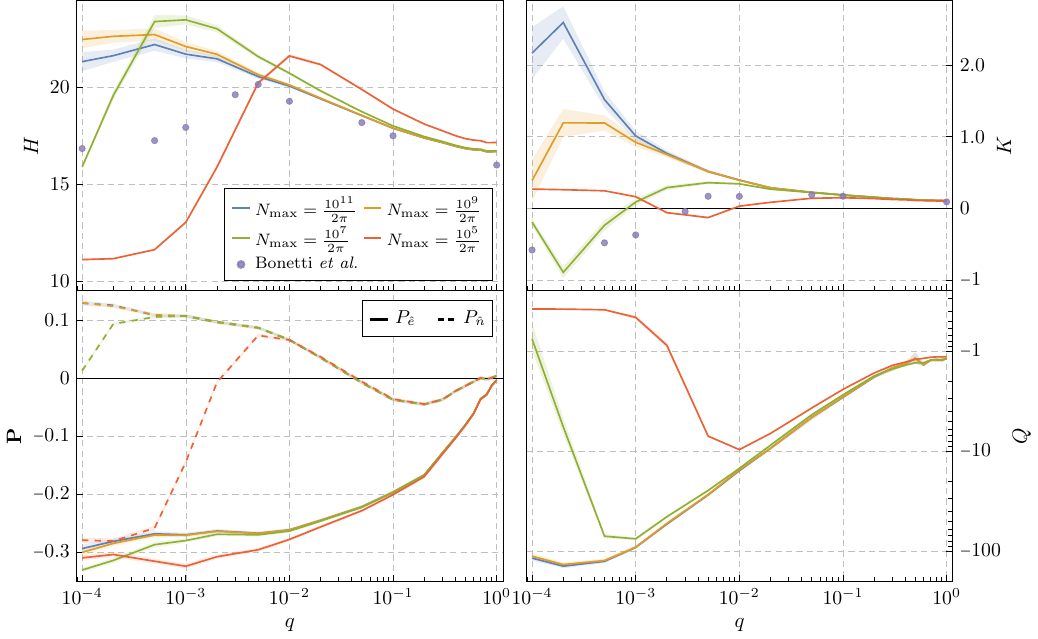}
\caption{Evolution parameters as functions of mass ratio $q$ for $\vec V = 0$ at fixed $e=0.6$ and $a/a_h=0.1$, for several choices of the integration-time cutoff $N\ped{max}$. The purple markers show the results digitized from Fig.~1 of Ref.~\cite{Bonetti:2020iwk}. For sufficiently small cutoffs, $K$ becomes negative at small $q$; increasing $N\ped{max}$ progressively restores the converged behavior with $K>0$.}
\label{fig:HK-q}
\end{figure*}

%%%%%%%%%%%%%%%%%%%%%%%%%%%%%%%%%%%%%%%%%%%%%%%%%%%%%%%%%%%
As described in Sec.~\ref{sec:num-int}, we classify an encounter as \emph{resolved} if the test particle leaves the sphere of radius $r\ped{sph}=50a$ with positive total energy and outward radial velocity, and as \emph{unresolved} if this does not occur within a prescribed time cutoff. Our default cutoff $N\ped{max} = 10^{11}/(2\pi)$ yields resolved fractions $\gtrsim 99\%$ across most of parameter space, with unresolved events concentrated at small $q$ and low incoming velocities (Tab.~\ref{tab:resolved-fractions}).
Figure~\ref{fig:HKPQ-Tmax} shows the evolution parameters as a function of $N\ped{max}$ for $a/a_h=0.1$, $e=0.6$ and $\vec V=0$, and several values of $q$. All parameters approach finite asymptotic values as $N\ped{max}$ is increased, but convergence is slower for smaller $q$, consistent with the $q^{-3/2}$ scaling of Eq.~\eqref{eq:Nesc}. In the regime $N\ped{max}\sim 10^4-10^5$, all parameters change dramatically. Truncating the simulations at shorter times would therefore yield qualitatively different results.
The hardening rate $H$ overshoots its asymptotic value before coming down as the longest encounters are included, an effect that reaches $\mathcal O(10\%)$ level for $q\ll1$. The precession parameter $Q$ is by far the most sensitive to long-lived interactions, with converged values differing by orders of magnitude from those at intermediate $N\ped{max}$. This is because $\Delta\varpi$ is accumulated via Eq.~\eqref{eqn:gauss} over the entire simulation, so a single long-lived event contributes far more precession than a short fly-by. The acceleration $\vec P$ also shows a noticeable cutoff dependence at small $q$, though less dramatic than $Q$. Most consequentially, $K$ exhibits a pronounced non-monotonic behavior at small $q$ as a function of $N\ped{max}$: it first decreases, passes through a negative minimum, and finally increases towards a positive asymptotic value.
Figure~\ref{fig:HK-q} offers a complementary view, plotting the evolution parameters as functions of $q$ at fixed $a/a_h$, for several values of $N\ped{max}$. For sufficiently small cutoffs, $K$ becomes negative at small $q$, while increasing $N\ped{max}$ progressively restores $K>0$. For the smallest $q$ shown, even $N\ped{max} = 10^{11}/(2\pi)$ may not be fully converged: features such as the turnover of $K$ at small $q$ continue to diminish as the cutoff is increased (see also Fig.~\ref{fig:bonetti-comparison} in Appendix~\ref{app:comparison}, where we ran a few simulations with longer integration times).

%%%%%%%%%%%%%%%%%%%%%%%%%%%%%%%%%%%
\subsection{Physical time cutoffs}
\label{sec:timescales}
%%%%%%%%%%%%%%%%%%%%%%%%%%%%%%%%%%%
While one would ideally let $T\ped{max}\to\infty$ and resolve all scatterings, there is both a computational and a physical reason for imposing a finite cutoff. The relevant question is thus how $T\ped{max}$ compares to the timescales that govern the binary's evolution. To answer it, we express all timescales in units of the binary's orbital period $T$. We will then evaluate these timescales in physical units for specific astrophysical systems in Sec.~\ref{sec:astro_impl}.

\medskip

\paragraph{Hardening timescale.} The most relevant timescale is the $e$-folding time of $1/a$,
\beq
T\ped{hard}=-\frac{a}{\dot a}=\frac{\sigma}{G\rho aH}\,.
\label{eq:T-hard}
\eeq
Particles that remain bound to the binary for longer than $T\ped{hard}$ interact with a binary whose semi-major axis has changed substantially. Their contribution to the scattering averages can then no longer be computed self-consistently. We can conveniently express this timescale through an approximation that is often applied to SMBH binaries (to which we return in Sec.~\ref{sec:smbhs}): we equate the influence radius $r_i=GM/\sigma^2$ with the radius within which the environment mass equals twice the binary mass, that is, $\frac43\pi r_i^3\rho\sim2M$. The number of orbits per hardening $e$-fold is then
\beq
N\ped{hard}\equiv\frac{T\ped{hard}}{T}\sim\frac{32}{3H}\,\frac{(1+q)^5}{q^{5/2}}\bigg(\frac{a_h}{a}\bigg)^{5/2}\,.
\label{eqn:Nhard}
\eeq
The ratio~\eqref{eqn:Nhard} depends only on the binary parameters and on $H$, and it grows rapidly for hard, asymmetric binaries. This cutoff typically lies well past the point where $K$ turns negative, as shown by vertical dashed lines in Fig.~\ref{fig:HKPQ-Tmax}.

\medskip

\paragraph{Hubble time.} An absolute upper bound on the length of any three-body interaction is the Hubble time, $T_H\sim\SI{e10}{yr}$. In units of the orbital period,
\beq
N_H\equiv\frac{T_H}{T}=\frac{4\sigma^3 T_H}{\pi G M}\,\frac{(1+q)^3}{q^{3/2}}\bigg(\frac{a_h}{a}\bigg)^{3/2}\,.
\label{eqn:Nmax-Hubble}
\eeq
For typical astrophysical systems, this exceeds $N\ped{hard}$ by a few orders of magnitude.

\medskip

\paragraph{GR precession timescale.} In addition to the three-body scattering-induced precession, the eccentricity vector also precesses due to general relativistic (GR) effects on a timescale $T\sped{gr}$. We can express this through a GR precession parameter $Q\sped{gr}$, whose definition mirrors that of $Q$ in \eqref{eqn:Q}. Its expression is
\beq
\begin{split}
Q\sped{gr}&=\frac{3\sigma(GM)^{3/2}}{G\rho c^2a^{7/2}(1-e^2)H}\\
&\sim\frac{256\pi}{(1-e^2)H}\frac{(1+q)^7}{q^{7/2}}\bigg(\frac{\sigma}c\bigg)^2\bigg(\frac{a}{a_h}\bigg)^{7/2}\,,
\end{split}
\label{eq:Q_gr}
\eeq
where the second line assumes again $G\rho\sim\frac{3}{2\pi}\sigma^6/(GM)^2$. In units of the binary period, the precession timescale then reads
\beq
N\sped{gr}=\frac{T\sped{gr}}T=\frac{(1-e^2)}{12}\frac{q}{(1+q)^2}\bigg(\frac{c}{\sigma}\bigg)^2\bigg(\frac{a_h}{a}\bigg)\,.
\eeq
GR precession is generally slower than the hardening timescale due to the large $(c/\sigma)^2$ factor, but $N\sped{gr}$ can become comparable or smaller than $N\ped{hard}$ for small $q$ and $a\ll a_h$.
In most cases of astrophysical interest, the hierarchy of timescales is thus $T\ll T\ped{hard}\ll T_H$ (equivalently, $1\ll N\ped{hard}\ll N_H$), with $T\sped{gr}$ potentially comparable to $T\ped{hard}$ for small $q$ or very hard binaries. The hardening timescale lies well within the regime where the evolution parameters have converged, as shown in Fig.~\ref{fig:HKPQ-Tmax}.
In contrast to earlier works, we find that $K$ remains positive across the full range of $q$ and $e$ explored. Three-body slingshots therefore drive eccentricity growth for all mass ratios, rather than circularizing the binary at small $q$. As we demonstrate explicitly in Appendix~\ref{app:comparison}, imposing the additional stopping criterion used in Refs.~\cite{Rasskazov_2019,Bonetti:2020iwk} reproduces their negative $K$, while removing it restores $K>0$. This confirms that the sign change is an artifact of truncating long-lived encounters. The implications of a universally positive $K$ for, e.g., the eccentricity distribution of PTA and LISA sources are discussed in Sec.~\ref{sec:astro_impl}.
%%%%%%%%%%%%%%%%%%%%%%%%%%%%%%%%%%%%%%%%%%%
\section{Solving for the binary evolution}
\label{sec:binary_evolution}
%%%%%%%%%%%%%%%%%%%%%%%%%%%%%%%%%%%%%%%%%%%
The simplest setting in which to study the time evolution of the binary parameters is an infinite uniform medium. With no density gradient, there is no restoring force from the environment, and the CoM force can operate unimpeded (in contrast to more realistic astrophysical scenarios, which we discuss in Sec.~\ref{sec:astro_impl}). This ``clean'' scenario allows us to study the secular evolution of the binary in detail, and elucidate the physics of the problem. In addition, it provides a testbed for validating our scattering results against direct $N$-body simulations, presented in Appendix~\ref{app:Nbody}.
%%%%%%%%%%%%%%%%%%%%%%%%%%%%%%%%%
\subsection{Equations of motion}
\label{sec:eoms}
%%%%%%%%%%%%%%%%%%%%%%%%%%%%%%%%%
We study the case of a binary initially at rest. As per the arguments in Sec.~\ref{sec:symmetry-rest} and~\ref{sec:motion-orbital-plane}, the binary's motion will remain confined to its orbital plane: the out-of-plane component of the CoM force vanishes, as does the rotation parameter. This case is actually rather generic, as a binary with an initially nonzero $V_{\hat L}$ will be subject to dynamical friction in the $\hat L$ direction, which quickly damps any out-of-plane motion, as can be seen in Fig.~\ref{fig:P-RQ-vsV}.
As in Sec.~\ref{sec:symmetry}, we keep the binary's mass $M$ and mass ratio $q$ constant throughout. The remaining binary parameters, instead, evolve according to the definitions of $K$, $\vec P$ and $Q$, which express the variation of the eccentricity, velocity and longitude of periapsis per logarithmic change of $a$. It is therefore natural to switch variable to $\log(a_h/a)$ and write the system of coupled evolution equations as
\begin{align}
\label{eqn:dedxi}
\frac{\dd e}{\dd\log(a_h/a)}&=K\,,\\
\label{eqn:dVdxi}
\frac1\sigma\frac{\dd\vec V}{\dd\log(a_h/a)}&=\vec P\,,\\
\label{eqn:dvarpidxi}
\frac{\dd\varpi}{\dd\log(a_h/a)}&=Q\,.
\end{align}
To determine the evolution as a function of time (rather than $a/a_h$) it suffices to solve the hardening equation, which we can write as
\beq
\label{eqn:dtdxi}
\frac{\dd t}{\dd\log(a_h/a)}=\mathcal T_h\frac{a_h/a}H\,,\qquad \mathcal T_h\equiv\frac{\sigma}{G\rho a_h}\,.
\eeq
The parameters $H$, $K$, $\vec P$, and $Q$ on the right-hand sides of equations \eqref{eqn:dtdxi}, \eqref{eqn:dedxi}, \eqref{eqn:dVdxi}, \eqref{eqn:dvarpidxi} are themselves functions of $a/a_h$, $e$, $\vec V$ and $\varpi$, as shown explicitly in Sec.~\ref{sec:results}. The $\varpi$-dependence amounts to a rotation of the binary axes $\hat e$ and $\hat n$ with respect to fixed axes $\hat x$ and $\hat y$. The dependences on $a/a_h$ and $\vec V$ are determined through the Maxwellian weighing described in Sec.~\ref{sec:extracting-evolution-parameters}. Finally, the $e$-dependence is obtained by interpolating scattering data computed on a fine grid with step size $\Delta e=0.01$. The propagation of uncertainties through the solution of these differential equations is rather involved, and we refer the reader to Appendix~\ref{app:uncertainties} for details.

\begin{figure*}
\centering
\includegraphics[width=\textwidth]{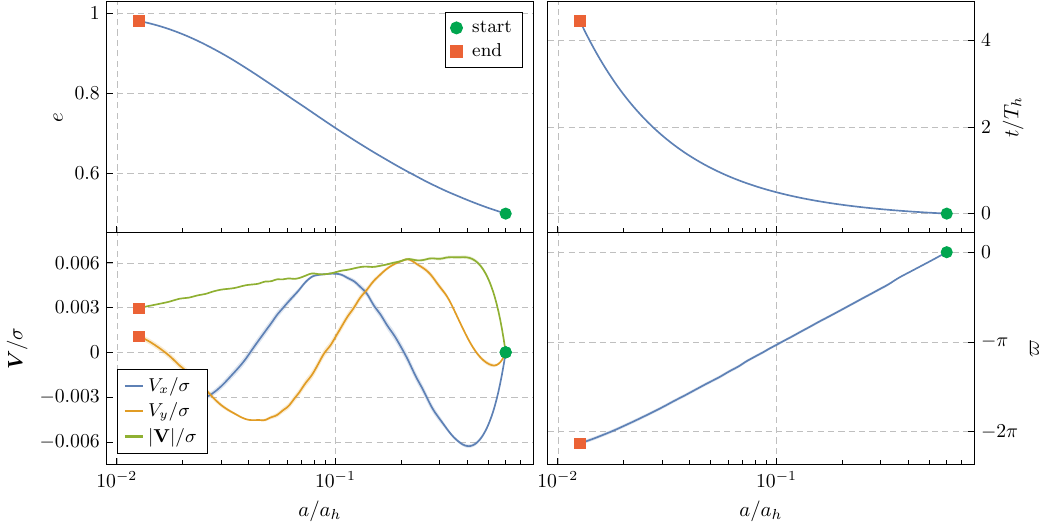}
\caption{Evolution of the binary parameters for $q=0.2$, obtained by solving equations \eqref{eqn:dedxi}, \eqref{eqn:dVdxi}, and~\eqref{eqn:dvarpidxi}. As an initial condition, we use $e=0.5$, $a/a_h=0.6$, and $\vec V=0$. The top-right panel shows the relation between time and binary hardness, as per Eq.~\eqref{eqn:dtdxi}. We assume $b\ped{max}=GM/\sigma^2$ in the evaluation of $\vec P\sped{df}$. The semi-transparent band around each line indicates the $1\sigma$ uncertainty.}
\label{fig:evolution}
\end{figure*}
\begin{figure}
\centering
\includegraphics[width=0.48\textwidth]{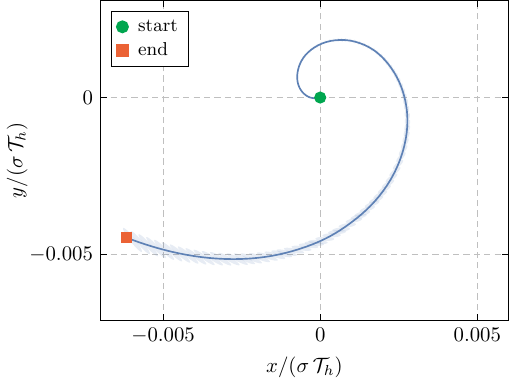}
\caption{Trajectory of the binary's CoM, for the same parameters and initial conditions as in Fig.~\ref{fig:evolution}. The semi-transparent ellipses denote the $1\sigma$ uncertainty at several points along the trajectory.}
\label{fig:trajectory}
\end{figure}

We show in Fig.~\ref{fig:evolution} a numerical solution to equations \eqref{eqn:dedxi}, \eqref{eqn:dVdxi}, \eqref{eqn:dvarpidxi}, and \eqref{eqn:dtdxi} for a binary with $q=0.2$, which is initially at rest, with $a/a_h=0.6$, and $e=0.5$ and aligned such that $\hat e=\hat x$ and $\hat n=\hat y$. We set $b\ped{max}=r_i=GM/\sigma^2$ in evaluating the dynamical friction contribution, $\vec P\sped{df}$, to $\vec P$, and halt the evolution when the binary reaches $e=0.98$.
The resulting motion of the binary's CoM can be described qualitatively as follows. The binary is initially at rest, so its CoM starts moving under the three-body scattering acceleration, $\vec P(0)=\vec P\sped{3bs}\ne0$. Very quickly, however, dynamical friction becomes important. From the linear approximation $\vec P\sped{df}\approx-\gamma\vec V/\sigma$ introduced in \eqref{eqn:P-small-V}, the CoM acceleration stalls at a terminal velocity
\beq
V\ped{term}=\frac{\sigma\,\abs{\vec P\sped{3bs}}}\gamma=\sigma\,\frac{3H}{16\sqrt{2\pi}}\frac{q\abs{\vec P\sped{3bs}}}{(1+q)^2}\frac{a}{a_h}\frac1{\braket{\log\Lambda}}\,.
\label{eqn:Vterm}
\eeq
This terminal velocity changes secularly: it decreases in magnitude (because $\gamma\propto a_h/a$) and rotates in direction (because the binary precesses), as shown in Fig.~\ref{fig:evolution}. The CoM therefore traces out an outward spiral trajectory, as shown in Fig.~\ref{fig:trajectory}. We have validated this behavior with direct $N$-body simulations of a binary immersed in a background of test particles, described in Appendix~\ref{app:Nbody}: as expected, the CoM velocity develops entirely in the orbital plane, and follows the same outward spiral (Fig.~\ref{fig:nbody-validation}).
Because $\abs{\vec P\sped{3bs}}\sim\mathcal O(1)$, Eq.~\eqref{eqn:Vterm} shows that the terminal velocity satisfies $V\ped{term}\ll\sigma$ for $a/a_h\ll1$. All the evolution parameters (besides $\vec P$) remain therefore very close to their $\vec V=0$ value throughout the evolution. The eccentricity evolution thus closely follows the results previously obtained in the literature, which ignored the variation of $\vec V$ and $\varpi$: the binary grows more eccentric as it hardens, with growth tapering off as $e\rightarrow1$ (which it never reaches) due to the decreasing $K$. The precession rate also remains approximately constant, due to the mild dependence of $Q$ on $a/a_h$, resulting in an approximately linear evolution of $\varpi$ over time (as can be seen in the bottom right panel of Fig.~\ref{fig:evolution}).
%%%%%%%%%%%%%%%%%%%%%%%%%%%%%%%
\subsection{Motion of the CoM}
\label{sec:trajectory}
%%%%%%%%%%%%%%%%%%%%%%%%%%%%%%%
In the small-$V$ limit, we can obtain an approximate analytical formula for the CoM trajectory by suppressing the dependence of the evolution parameters on $a/a_h$ and $e$, except for the dynamical friction term $\vec P\sped{df}\approx-\gamma\vec V/\sigma$, for which we assume $\gamma=C(a_h/a)\gg1$ with
\beq
C=\frac{16\sqrt{2\pi}}{3H}\frac{(1+q)^2}q\braket{\log\Lambda}
\eeq
a constant. In this limit, the equations \eqref{eqn:dVdxi}, \eqref{eqn:dvarpidxi}, and \eqref{eqn:dtdxi} can be solved with the initial conditions $a=a_0$ and $t=\varpi=\vec V=0$ to give
\begin{align}
\label{eqn:t-exact}
t&=\frac{\mathcal T_h}{H}\bigg(\frac{a_h}a-\frac{a_h}{a_0}\bigg)\,,\\
\label{eqn:varpi-exact}
\varpi&=Q\log(a_0/a)\,,\\
\label{eqn:V-exact}
\!\frac{\vec V}\sigma&=\bigg(\frac{\mathsf{R}(\varpi)}{C(a_h/a)}+\frac{\mathsf{R}(\varpi)(\mathsf{I}-Q\,\mathsf{R}(\frac\pi2))}{C^2(a_h/a)^2}+\ldots\bigg)\vec P\sped{3bs}\,,
\end{align}
where $\mathsf{I}$ is the identity matrix and $\mathsf R(\theta)$ is the rotation matrix of angle $\theta$ about $\hat L$. We can then solve for the CoM position $\vec r$ as
\beq
\vec r(t)=\frac{\mathcal T_h\sigma}{HC}\mathsf R(\varpi)\bigg(\frac{\mathsf{R}(-\frac\pi2)}{Q}-\frac{\mathsf{I}}{C(a_h/a)}+\ldots\bigg)\vec P\sped{3bs}\,,
\eeq
where we have omitted an integration constant $\vec r(0)$. The two terms in parentheses are orthogonal to each other. The prefactor and the first term have a weak dependence on $a/a_h$ (through $Q$), while the second term rapidly decays away as the binary hardens. The CoM thus traces out a circling trajectory whose radius slowly evolves over time, thus describing a spiral. The angular velocity of the CoM motion decreases as $\sim1/t$, according to \eqref{eqn:t-exact} and \eqref{eqn:varpi-exact}. At late times, when the binary has hardened and $C(a_h/a)\gg1$, the CoM undergoes an approximate circular motion, with radius given by
\beq
r=\frac{\sigma \mathcal T_h\abs{\vec P\sped{3bs}}}{HC\abs{Q}}=\frac{3\sigma^2\abs{\vec P\sped{3bs}}}{16\sqrt{2\pi}G\rho a_h\abs{Q}}\frac{q}{(1+q)^2}\frac1{\braket{\log\Lambda}}\,.
\label{eqn:circle-radius}
\eeq
For the binary parameters of Figs.~\ref{fig:evolution} and~\ref{fig:trajectory}, at $a/a_h=0.1$, we have $\abs{\vec P\sped{3bs}/Q}\sim0.12$ and $HC\approx96\braket{\log\Lambda}$. For $\braket{\log\Lambda}=0.673$, this gives a radius of $r\sim2\times10^{-3}\sigma \mathcal T_h$, in excellent agreement with Fig.~\ref{fig:trajectory}. We express this radius in terms of physical scales and discuss its astrophysical implications in Sec.~\ref{sec:astro_impl}.
Finally, we note that our choice $b\ped{max}=r_i=GM/\sigma^2$ is consistent (up to $\mathcal O(1)$ factors) with the resulting motion of the CoM. More specifically, formula \eqref{eqn:Fdf-V} for dynamical friction assumes that the binary moves in a straight line. But from \eqref{eqn:t-exact} and \eqref{eqn:varpi-exact} we see that the velocity direction changes by an angle of $\mathcal O(1)$ in a time $\sim \mathcal T_h/H$. Considering that a distant encounter with a particle with impact parameter $b$ lasts for a time $\sim b/\sigma$, we find that our treatment of dynamical friction is consistent when
\beq
\frac{b}\sigma\lesssim\frac{\mathcal T_h}H\implies b\lesssim\frac{8\pi}{3H}\frac{(1+q)^2}qr_i\,.
\eeq
For the binary parameters used in Figs.~\ref{fig:evolution} and~\ref{fig:trajectory}, for which $H\sim20$, this gives $b\lesssim3r_i$.
%%%%%%%%%%%%%%%%%%%%%%%%%%%%%%%%%%%%%
\section{Astrophysical implications}
\label{sec:astro_impl}
%%%%%%%%%%%%%%%%%%%%%%%%%%%%%%%%%%%%%
In the previous sections, we studied the problem in a deliberately clean setup: a binary embedded in a uniform, collisionless medium, without fixing the absolute mass or density scale. This allowed us to identify new effects such as a nonzero CoM force and apsidal precession. We now explore several astrophysical implications of these results. For each scenario, we choose a physical scale for the binary and its environment, and evaluate all time and length scales in physical units.
We do not expect the discussion in this section to be exhaustive. We limit ourselves to a qualitative overview of the implications that we deem most interesting. Due to the diverse set of scenarios involved, each of them (and possibly more) deserves a dedicated future investigation.
%%%%%%%%%%%%%%%%%%%%%%%%%%%%%%%%%%%%%%%%%%%%%%%%%%%%%%%%%%%
\subsection{Stellar hardening of supermassive black holes}
\label{sec:smbhs}
%%%%%%%%%%%%%%%%%%%%%%%%%%%%%%%%%%%%%%%%%%%%%%%%%%%%%%%%%%%
Perhaps the most immediate application of our results is to SMBH binaries hardening in stellar backgrounds. This is indeed the scenario that motivated the earliest works on the topic \cite{Quinlan:1996vp,Sesana:2006xw,Sesana:2006ne,Sesana:2007vr}. Our findings, such as the CoM force and the consequent binary motion, are relevant for several aspects of stellar hardening previously studied in the literature. These include the SMBH Brownian motion \cite{Merritt:2000yg,Merritt:2001hy,Milosavljevic:2002bn,Chatterjee:2002sq,Chatterjee:2003ji,Berczik:2005ff,Bortolas:2016lge,Khan:2020,Varisco_2021}, and the so-called \emph{final parsec problem} \cite{Quinlan:1997qe,Milosavljevic:2001vi,Milosavljevic:2002bn,Makino:2003da,Berczik:2005ff}, that is, the question of whether the supply of stars to the SMBHs is enough to bring the binary to merger within a Hubble time. After evaluating the relevant length and timescales in physical units, we provide a discussion of those problems.
%%%%%%%%%%%%%%%%%%%%%%%%%%%%%%%%
\subsubsection{Physical scales}
%%%%%%%%%%%%%%%%%%%%%%%%%%%%%%%%
We make use of the empirical relation between the mass $M$ of a SMBH and the velocity dispersion $\sigma$ of the stars around it, known as $M$--$\sigma$ relation.\footnote{We assume here that the $M$--$\sigma$ relation still holds for a SMBH binary, with $M$ its total mass.} We use the parameters from~\cite{Kormendy:2013dxa}, according to which the relation takes the form
\beq
M\approx10^{8.49}M_\odot\bigg(\frac{\sigma}{\SI{200}{km/s}}\bigg)^{4.38}\,.
\label{eqn:M-sigma}
\eeq
The formation, evolution and merger of SMBH binaries is a complex process involving distinct phases \cite{Merritt:2004gc}. When two galaxies merge, the respective SMBHs sink towards the center of the newly formed galaxy under the effect of dynamical friction, in a timescale of $\num{e6}-\SI{e9}{yr}$ \cite{Begelman:1980vb,Dosopoulou:2016hbg}. This process stops being effective when the two BHs form a hard binary, i.e., when they come within a distance of order $a_h$ from each other. Using \eqref{eqn:M-sigma}, we can express the hard binary radius, defined in \eqref{eqn:a_h}, as
\beq
a_h=\SI{1.29}{pc}\,\bigg(\frac{M}{10^7M_\odot}\bigg)^{0.54}\frac{q}{(1+q)^2}\,.
\eeq
At $a\lesssim a_h$, the binary hardens primarily through scatterings of stars, precisely of the type studied in the preceding sections. This process continues until the separation falls to about
\beq
\begin{split}
a_g&\equiv\frac{GM}{c\sigma}\bigg(\frac\mu{M}\bigg)^{1/5}\\
&=\SI{1.6e-3}{pc}\,\bigg(\frac{M}{10^7M_\odot}\bigg)^{0.77}\frac{q^{1/5}}{(1+q)^{2/5}}\,,
\end{split}
\eeq
below which gravitational-wave emission takes over and drives the binary to merger.
It is also useful to have at hand physical timescales, some of which we have already discussed in Sec.~\ref{sec:timescales}. The binary's orbital period reads
\beq
T=\SI{4.3e4}{yr}\,\frac{q^{3/2}}{(1+q)^3}\bigg(\frac{M}{10^7M_\odot}\bigg)^{0.32}\bigg(\frac{a}{a_h}\bigg)^{3/2}\,.
\eeq
The hardening timescale, under the assumption that $r_i=GM/\sigma^2$ satisfies $\frac43\pi r_i^3\rho\sim2M$ (an approximation introduced in Sec.~\ref{sec:timescales}), is
\beq
T\ped{hard}\sim\SI{2.3e4}{yr}\bigg(\frac{20}H\bigg)\frac{(1+q)^2}{q}\bigg(\frac{M}{10^7M_\odot}\bigg)^{0.32}\bigg(\frac{a_h}{a}\bigg)\,.
\eeq
This is always much longer than $T$ for $a\ll a_h$, and shorter than the Hubble time $T_H=\SI{e10}{yr}$ in most of the parameter space. The evolution parameters derived in Secs.~\ref{sec:results} and~\ref{sec:Tmax} for $T\ped{max}\to\infty$ can thus be safely applied to SMBH binaries.
The GR precession parameter reads
\beq
Q\sped{gr}\sim\frac{\num{3.7e-6}}{1-e^2}\bigg(\frac{20}H\bigg)\frac{(1+q)^7}{q^{7/2}}\bigg(\!\frac{M}{10^7M_\odot}\!\bigg)^{\!0.46}\bigg(\frac{a_h}a\bigg)^{\!7/2}\,.
\eeq
This should be compared to $Q$, which has the opposite sign, is typically of $\mathcal O(1)$ for moderate mass ratios, and roughly increases as $q^{-2/3}$ for small $q$, as can be read off Fig.~\ref{fig:HKPQ-Tmax}. It is thus also safe to assume that, for all but the very hard binaries with $a\lesssim10^{-2}a_h$, the slingshot-induced precession dominates over the GR one for SMBH binaries.
%%%%%%%%%%%%%%%%%%%%%%%%%%%%%%%%%%%%%%%%%%%%%%%%%%%%%%
\subsubsection{Brownian motion and terminal velocity}
%%%%%%%%%%%%%%%%%%%%%%%%%%%%%%%%%%%%%%%%%%%%%%%%%%%%%%
As stated in Sec.~\ref{sec:binary-uniform-medium}, all our results are valid in the ``smooth medium'' limit, that is, $m\to0$, $n\to\infty$, with $\rho=mn$ constant. A granular medium, composed of stars whose mass is $m\ne0$, introduces stochasticity in the evolution of the binary \cite{Merritt:2000yg}. One manifestation is a Brownian motion of the binary's CoM, whose typical velocity follows from energy equipartition:
\beq
V\ped{Brown}=\sigma\sqrt{\frac{m}M}=\num{3e-4}\,\sigma\,\bigg(\frac{10^7m}{M}\bigg)^{1/2}\,.
\label{eqn:V-Brown}
\eeq
This should be compared with the terminal velocity arising from the equilibrium of the CoM force and dynamical friction. From \eqref{eqn:Vterm}, we have
\beq
V\ped{term}=0.15\,\sigma\bigg(\frac{H}{20}\bigg)\frac{q\abs{\vec P\sped{3bs}}}{(1+q)^2}\bigg(\frac{a/a_h}{0.1}\bigg)\frac1{\braket{\log\Lambda}}\,.
\label{eqn:V-term-typical}
\eeq
We thus see that, for typical values of the parameters such as the ones chosen in \eqref{eqn:V-Brown} and \eqref{eqn:V-term-typical}, the nonzero terminal velocity caused by the CoM force is larger than the typical SMBH Brownian motion. This remains true even for relatively small mass ratios, such as $q=0.2$ chosen in Fig.~\ref{fig:evolution}.
We thus reach a striking conclusion: during the stellar hardening phase, the motion of a typical SMBH binary's CoM is expected to be primarily smooth and deterministic, following the spiral (or approximately circular) trajectory shown in Fig.~\ref{fig:trajectory}. Brownian noise is a small perturbation on top of this motion, becoming more relevant as the binary hardens, owing to the decreasing $V\ped{term}$.
The binary's orientation also undergoes a so-called rotational Brownian motion \cite{Merritt:2001hy}. This effect reorients the binary by an angle of order $\sqrt{m/M}$ during the time its semi-major axis shrinks by an $e$-fold. The average, smooth precession we discussed in earlier sections induces a rotation of the binary axes $\hat e$ and $\hat n$ by an angle $Q\sim\mathcal O(1)$ per hardening $e$-fold. In this case too, we thus conclude that average smooth precession is dominant over the stochastic Brownian component.
The other way the binary can reorient is through a rotation of its orbital plane. We did not study this in Sec.~\ref{sec:binary_evolution}, because that is only switched on when $V_{\hat L}\ne0$, and dynamical friction is expected to bring $V_{\hat L}$ to vanish. However, Brownian velocity $V_{\hat L}\sim\sigma\sqrt{m/M}$ can activate the mechanism. Assuming for the rotation parameter $\abs{\vec R}\sim V/\sigma$ (cf.\ Fig.~\ref{fig:P-RQ-vsV}), we find that the orbital plane reorients by an angle of order $\sqrt{m/M}$ per hardening $e$-fold. This rotation depends deterministically on the CoM Brownian motion, and comes on top of the purely rotational Brownian motion cited above. The two effects are of the same order.
%%%%%%%%%%%%%%%%%%%%%%%%%%%%%%%%%%%%%%%%%%%%%%%%%%%%%%%%
\subsubsection{CoM motion and the final parsec problem}
\label{sec:final-parsec}
%%%%%%%%%%%%%%%%%%%%%%%%%%%%%%%%%%%%%%%%%%%%%%%%%%%%%%%%
Efficient stellar hardening of a SMBH binary relies on a continuous supply of stars on low-angular-momentum orbits, which come close enough to the binary to participate in a three-body interaction and be slingshotted away. Because these stars are ejected after one such interaction, the rate of three-body scatterings is dictated by how efficiently \emph{other} stars can diffuse into this low-angular-momentum region of phase space, known as the \emph{loss cone}. Galactic two-body relaxation timescales are too long to sustain efficient binary hardening and bring the SMBH separation from $a_h$ to $a_g$ (and hence to merger) within a Hubble time. This puzzle is known as the final parsec problem.
A great deal of work has investigated which physical processes could refill the loss cone efficiently enough to allow SMBHs to merge. A leading candidate is collisionless loss cone refilling \cite{Gualandris:2017,Khan:2011bh,Preto:2011gu}, i.e., angular momentum diffusion induced by non-sphericity of the galaxy. While this idea has led to a broad consensus that the final parsec problem is unlikely to be a genuine issue in realistic galaxies, the coalescence timescale of SMBH binaries is still highly uncertain and depends on several poorly constrained aspects.
Understanding how the CoM motion affects the loss cone refilling is thus interesting for at least two reasons: (i) it could efficiently replenish the loss cone even for spherical galaxies, and (ii) understanding all contributions to the binary's dynamics is essential to make reliable predictions of merger rates and GW signals.
The connection between CoM motion (either Brownian or induced by rotation of the stellar cluster) and loss cone refilling has indeed been studied in the literature \cite{Milosavljevic:2002bn,Varisco_2021}. It is then natural to ask how these works should be revised or extended in light of our findings. Although we postpone a detailed analysis to a future work, we provide here a qualitative discussion of the most likely relevant aspects.
By moving around, rather than sitting at the origin, the binary is able to access a larger portion of the stellar phase space. According to our discussion in Sec.~\ref{sec:trajectory}, the binary's CoM is expected to move along a spiral that, at late times, approaches a circle when the dependence of $\abs{\vec P\sped{3bs}}$, $Q$ and $C$ on the binary parameters is neglected. The radius of such circle is given in \eqref{eqn:circle-radius} as a function of $\sigma$, $\rho$, and $a_h$. To express it in physical units, we once again approximate $G\rho\sim\frac{3}{2\pi}\sigma^6/(GM)^2$. The radius of the circle then becomes
\beq
r\sim\frac{\sqrt{2\pi}}{4}\frac{\abs{\vec P\sped{3bs}}}{\abs{Q}}\frac1{\braket{\log\Lambda}}r_i\,.
\eeq
Apart from a prefactor of order $\mathcal O(0.1)$, the binary's CoM trajectory spans a physical distance comparable to its own influence radius,
\beq
r_i= \SI{5.2}{pc}\,\bigg(\frac{M}{10^7M_\odot}\bigg)^{0.54}\,.
\eeq
For hard binaries with $a\ll a_h$, this is much larger than the semi-major axis $a$.
To have a chance to undergo a three-body interaction followed by a slingshot, stars then only need to come within a distance $r_i$ from the galactic center, rather than the much smaller $a$. This naturally enlarges the size of the loss cone, and thus the number of stars initially available to the SMBH binary, lifting the critical angular momentum from $L_{c,1}\sim\sqrt{GMm^2a}$ to $L_{c,2}\sim\sqrt{GMm^2r_i}$. The fraction of stars within a distance $r_i$ from the binary that is in the loss cone thus dramatically increases from $L_{c,1}^2/L_0^2\sim a/r_i\ll1$ to $L_{c,2}^2/L_0^2\sim1$, where $L_0=m\sigma r_i$ is the circular angular momentum at $r_i$. Essentially \emph{all} the stars in the binary's radius of influence can undergo a three-body interaction, and help the binary harden further. The total stellar mass within $r_i$ is $\sim2M$. Because each scattered star carries away an energy $\sim GMm/a$, if the binary interacts with the entire cluster, the total energy lost is $\sim GM^2/a$, which is comparable to the binary's binding energy.
The above argument is a simple dimensional estimate, but it demonstrates that a CoM motion of order $r_i$ is precisely what is needed for the binary to lose the entirety of its energy, and reach the GW-dominated phase.
There are several details and complications that could be taken into account, and quantified, in a more detailed study. The first of these is that the binary's motion itself contributes to the diffusion of stars in angular momentum space, thus further helping to refill the enlarged loss cone. This makes the case for the solution, or at least alleviation, of the final parsec problem even stronger. On the other hand, a critical element to take into account in an accurate analysis is the gravitational potential the SMBH binary sits in. In our discussion in Sec.~\ref{sec:binary_evolution}, we considered a uniform medium, with no density gradient nor restoring force. In reality, the gravity of the nuclear cluster will confine the motion of the binary's CoM. Including this potential in the analysis is, however, not as trivial as adding a static background potential. The reason is that the binary itself induces a strong feedback on the inner region of the cluster, which, for example, undergoes core scouring and can no longer remain spherically symmetric due to the binary's motion in it.
As a final observation, we note that most proposed solutions to the final parsec problem involve an ``external'' or ``environmental'' element that diffuses stars into the loss cone. Given the surprisingly rich physics of the stellar three-body scattering we found in this paper, it will be particularly appealing if stellar scattering can, after all, take care of its own supply, potentially eliminating the final parsec problem at its root.
%%%%%%%%%%%%%%%%%%%%%%%%%%%
\subsubsection{GW signals}
%%%%%%%%%%%%%%%%%%%%%%%%%%%
A population of inspiraling SMBH binaries is believed to source the stochastic GW background recently detected by Pulsar Timing Arrays (PTA) \cite{NANOGrav:2023gor,EPTA:2023fyk,Reardon:2023gzh,Xu:2023wog,NANOGrav:2023hfp,EPTA:2023xxk}. The low-frequency portion of the PTA spectrum is sensitive to environmental effects, such as the interaction of the binary with its stellar background. This part of the signal has indeed been modeled using inputs from the three-body scattering literature \cite{Chen:2024bbg}, in particular the values of $H$ and $K$.
Our work is relevant for analyses of the PTA signal in at least two ways. First, compared to previous studies, our suite of three-body scattering experiments is more extensive in terms of number of simulations per pair of $(q,e)$, as well as in terms of parameter space coverage. In addition, $H$ and $K$ receive a small correction due to the binary's CoM motion, which we calculate for the first time. Updating the existing studies with our numerical data will be straightforward.
A second way our results can influence the interpretation of the stochastic GW background is more subtle. If, as speculated in Sec.~\ref{sec:final-parsec}, the CoM displacement plays a significant role in refilling the loss cone, then the hardening of symmetric binaries with $q\approx1$ or $e\approx0$ will be slower than that of asymmetric binaries, due to the suppression of the CoM force. The parameters of the SMBH binary population seen by PTA will thus be biased (compared to those inferred from SMBH mass functions) because the evolution of equal-mass binaries stalls at low frequencies. It would be interesting to re-analyze the PTA signal using appropriately frequency-dependent parameters for the SMBH binary population.
%%%%%%%%%%%%%%%%%%%%%%%%%%%%%%%%%%%%%%%%%%%%%
\subsection{Beyond supermassive black hole binaries}
%%%%%%%%%%%%%%%%%%%%%%%%%%%%%%%%%%%%%%%%%%%%%
Although previous work on stellar hardening was primarily motivated by SMBH binaries in galactic nuclei, the evolution parameters $H$, $K$, $\vec P$, $Q$, and $\vec R$ derived in Sec.~\ref{sec:results} apply more broadly to any hard binary interacting with a background of much lighter particles. In this section we briefly discuss other astrophysical settings in which the new effects identified here may be relevant.
%%%%%%%%%%%%%%%%%%%%%%%%%%%%%%%%%%%%%%%%%%
\subsubsection{Small-mass-ratio binaries} 
%%%%%%%%%%%%%%%%%%%%%%%%%%%%%%%%%%%%%%%%%%
Intermediate mass-ratio inspirals (IMRIs, $q\sim10^{-2}-10^{-4}$) and extreme mass-ratio inspirals (EMRIs, $q\lesssim10^{-4}$) are prime targets for LISA \cite{LISA:2024hlh}. These systems combine small mass ratios with potentially large eccentricities, placing them in the regime where our new results are most pronounced: the acceleration parameter $\vec P$ and the precession parameter $Q$ are largest (Fig.~\ref{fig:P_xy-Q}), and the sign of the eccentricity growth rate $K$ is different from previous claims in the literature.
Specifically, as established in Sec.~\ref{sec:Tmax} and Appendix~\ref{app:comparison}, $K$ remains positive across the full range of mass ratios explored, in contrast with the $K<0$ at small $q$ reported in Refs. \cite{Rasskazov_2019,Bonetti:2020iwk}. To give an order-of-magnitude estimate, at $q = 0.001$, $e = 0.6$, and $a/a_h = 0.1$, we find $K\approx 1.0$ (Fig.~\ref{fig:HK-q}), meaning the eccentricity increases by order-unity per $e$-fold of binary hardening (before saturating as $e\rightarrow 1$).
This impacts the eccentricity distribution of small-mass-ratio binaries with potentially observable consequences. EMRIs formed through two-body relaxation are expected to have extremely high eccentricities \cite{Babak:2017tow,Mancieri:2025cmx}. GW emission subsequently circularizes these orbits, though a fraction of them is still expected to enter the LISA band with moderate-to-high eccentricities ($e \sim \mathcal{O}(0.1)$). During the early inspiral, where the GW and two-body relaxation timescales are comparable, eccentricity growth from stellar scattering could, in principle, counteract this circularization. This would systematically raise the eccentricity with which binaries enter the LISA band, thereby also shortening their GW-driven inspiral \cite{Peters:1963ux,Peters:1964zz}.
On the other hand, although $\vec P\sped{3bs}$ and $Q$ are largest for small $q$ and high $e$ (Figs.~\ref{fig:P_xy-Q} and~\ref{fig:orbit-grid}), both effects are subdominant to competing effects in the small-$q$ regime. For an IMBH-SMBH binary with $q=0.01$, $e=0.6$, and $a/a_h=0.1$, we have $|\vec P\sped{3bs}|\approx 0.2$ and $H\approx20$, giving a terminal velocity $V_\mathrm{term}\sim\num{4e-4}\sigma$ from \eqref{eqn:V-term-typical}. The Brownian velocity for a $10^6M_\odot$ SMBH surrounded by solar-mass stars is $V_\mathrm{Brown}\sim10^{-3}\sigma$, a factor of a few larger. This suppression arises because, at fixed $a/a_h$, the terminal velocity scales as $q|\vec P\sped{3bs}|/(1+q)^2$~\eqref{eqn:Vterm}, which decreases faster with $q$ than $|\vec P\sped{3bs}|$ grows. The CoM Brownian motion is known to produce Doppler and aberrational phase shifts in the emitted gravitational waveform~\cite{Torres-Orjuela:2025jtq}. While it is unlikely that the correction due to the CoM force will produce detectable GW signatures, we note that its imprint would likely take a qualitatively distinct form compared to the stochastic component.

Similarly, although the precession parameter $Q$ is largest for smaller mass ratios, reaching values of order $10-100$ (Fig.~\ref{fig:HK-q}), relativistic precession dominates in this regime. The GR precession $Q\sped{gr}$ scales as $q^{-7/2}$~\eqref{eq:Q_gr}, which grows much faster than $|Q| \propto q^{-2/3}$ as $q \rightarrow 0$. For all but the widest binaries, the three-body precession is therefore subdominant.
%%%%%%%%%%%%%%%%%%%%%%%%%%%%%%%%%%%%%%%%%%%%%%%%%
\subsubsection{Binaries in gaseous environments}
%%%%%%%%%%%%%%%%%%%%%%%%%%%%%%%%%%%%%%%%%%%%%%%%%
The symmetry argument underlying the CoM force and apsidal precession is not restricted to stellar backgrounds. Any asymmetric binary moving through a medium will, in general, scatter that medium anisotropically and recoil. This suggests that analogous secular effects may operate for binaries embedded in gas, such as stellar-mass BH binaries in active galactic nuclei (AGN) disks and EMRIs in accretion disks.
Applying our results quantitatively in that context requires caution. AGN disks are gaseous, flattened, and shearing, whereas our calculations assume a homogeneous, isotropic background of collisionless particles. Still, for a gas parcel at a distance $r\sim a$ from the binary, the gravitational acceleration $a\ped{grav}$ exerted by the binary dominates over the pressure-gradient acceleration $a\ped{press}$,
\beq
\frac{a\ped{grav}}{a\ped{press}} \sim \frac{GM/a^2}{c_s^2/a} \sim \frac{r_i}{a}\,,
\eeq
which is large in the hard-binary regime $a\ll r_i$, assuming $\sigma\sim c_s$. The slingshot mechanism responsible for $\vec P$ and $Q$ should therefore survive, at least qualitatively, in gaseous media. The main uncertainties are then associated with the disk geometry,\footnote{In principle, our scattering experiments could also be repeated for a planar background to better capture the disk geometry.} together with gas-specific effects such as pressure gradients, shocks, and radiative cooling, all of which may modify the effective evolution parameters.
A useful concrete example is provided by stellar-mass BH binaries near AGN migration traps, which have attracted considerable attention as candidate progenitors of LIGO/Virgo/KAGRA events \cite{Bartos:2016dgn,Stone:2016wzz,Leigh:2017wff,Mckernan:2017ssq,Secunda:2018kar,Tagawa:2019osr,Grobner:2020drr}. To give a sense of the relevant physical scales, we consider a $60M_\odot$ binary with mass ratio $q=0.3$ near a migration trap \cite{Bellovary:2015ifg} in a Sirko-Goodman disk \cite{Sirko:2002ex} around an SMBH of mass $M_\bullet=10^8M_\odot$, with accretion rate $\dot M_\bullet=0.5\dot M\ped{Edd}$ and viscosity parameter $\alpha=0.01$. At the trap location $R\ped{trap}\sim 10^3GM_\bullet/c^2$, the midplane density and sound speed are $\rho\sim10^{-9}\mathrm{g/cm^{3}}$ and $c_s\sim231\mathrm{km/s}$ \cite{Secunda:2018kar,Gangardt:2024bic}, respectively. Identifying $\sigma\sim c_s$, the hard-binary scale~\eqref{eqn:a_h} becomes
\beq
a_h \sim \SI{0.25}{AU} \bigg(\frac{M}{60 M_\odot}\bigg) \frac{q}{(1+q)^2} \bigg(\frac{\SI{231}{km/s}}{c_s}\bigg)^{2}\,.
\eeq
At the trap radius, the disk aspect ratio is $H\ped{d}/r \sim 0.017$, giving a scale height $H\ped{d}\sim 17\,\mathrm{AU}$. Thus, for hard binaries, $a\ll H\ped{d}$, so the interaction with the surrounding gas is approximately local and three-dimensional, although both $c_s$ and $H\ped{d}$ are uncertain at the order-of-magnitude level and depend on the adopted disk model.
The hardening timescale~\eqref{eq:T-hard} for such binaries is then
\beq
\begin{aligned}
T\ped{hard} \sim{}& \SI{1.4e3}{yr}\, \bigg(\frac{20}{H}\bigg) \bigg(\frac{10^{-9}\mathrm{g/cm^3}}{\rho}\bigg) \\&\times\bigg(\frac{c_s}{\SI{231}{km/s}}\bigg)\bigg(\frac{0.25\mathrm{AU}}{a_h}\bigg)\bigg(\frac{0.1}{a/a_h}\bigg)\,,
\end{aligned}
\label{eq:hardening_AGN}
\eeq
much shorter than a typical AGN lifetime of $\sim10^7-10^8\,\mathrm{yr}$. To the extent that the three-body picture remains valid in a gaseous medium, binaries in this regime can therefore undergo several hardening $e$-folds through repeated interactions with the disk gas, and the new effects identified in this work have ample time to leave an imprint.
The CoM motion follows the same logic as in the stellar case. In the subsonic regime ($V\ll c_s$), the gaseous dynamical-friction force takes the same parametric form as Chandrasekhar drag \cite{1999ApJ...513..252O} with $c_s$ replacing $\sigma$, and an effective Coulomb logarithm $\langle\log\Lambda\rangle\sim 3$ as suggested by simulations \cite{2013MNRAS.429.3114C}. Combined with $|\vec P\sped{3bs}|\sim 0.08$ (for $e=0.3$), Eq.~\eqref{eqn:Vterm} yields a terminal drift speed $V\ped{term}\sim 5.6\times10^{-4} c_s \sim \SI{131}{m/s}$. The corresponding CoM excursion is of order $\mathcal{O}(10\,\mathrm{AU})$. This drift could be dynamically important within the migration trap depending on the radial profile of the migration torque.
%%%%%%%%%%%%%%%%%%%%%%%%%%%%%%%%%%%
\subsubsection{Other environments}
%%%%%%%%%%%%%%%%%%%%%%%%%%%%%%%%%%%
In a similar fashion, our results can be applied to other environments. For example, a massive binary evolving within a dark matter spike \cite{Gondolo:1999ef} would experience the same self-acceleration and precession. Whether the dark matter density is ever large enough for these effects to be dynamically important depends on the spike's profile, which remains uncertain \cite{Merritt:2002vj,Sharpe:2026nqq}. Wide binary stars evolving through dense molecular clouds would also experience similar effects. Finally, we note that the restricted three-body approximation ($m\ll M$) underpins all our results. For binaries in environments where the background particle mass is not negligible compared to the binary mass---such as stellar-mass black hole binaries in globular clusters, where $m/M \sim 0.1$---corrections at the $\mathcal{O}(m/M)$ level are expected, and a full three-body treatment would be needed to capture them.
%%%%%%%%%%%%%%%%%%%%%%%%
\section{Conclusions}
\label{sec:conclusions}
%%%%%%%%%%%%%%%%%%%%%%%%
In this work, we have revisited the well-known problem of binary hardening in a background of lighter particles. We have shown that the problem is substantially richer than previously appreciated. Earlier treatments considered only the evolution of semi-major axis and eccentricity, expressed in terms of the hardening rate $H$ and eccentricity growth rate $K$. By combining symmetry arguments with an extensive suite of three-body scattering experiments, we identify three additional, generically nonvanishing effects: a net force on the binary's CoM, an apsidal precession, and---when the CoM has an out-of-plane velocity---a rotation of the orbital plane. We encode these in three new dimensionless parameters, $\vec P$, $Q$, and $\vec R$, which together with $H$ and $K$ form a complete framework for the secular evolution of hardening binaries.
The physical origin of these effects is simple but has not previously been recognized: because an asymmetric binary scatters background particles anisotropically, the recoil does not average to zero. As a result, even in a perfectly uniform and isotropic medium, the binary \emph{self-accelerates}. Only circular and equal-mass binaries are exempt, by virtue of their enhanced symmetry. For binaries at rest with respect to the background, the self-acceleration (in units of the velocity dispersion $\sigma$) and precession are both generically of order unity per hardening $e$-fold. Once dynamical friction on the binary as a whole is included, the CoM settles into a terminal velocity, whose direction is continually rotated by the slingshot-induced precession, producing an outward-spiraling trajectory in the orbital plane. We validate this prediction with direct $N$-body simulations. For SMBH binaries, the characteristic radius of this motion can be comparable to the binary’s radius of influence $r_i$, implying that hardening SMBH binaries should not, in general, be expected to sit at the centers of their host galaxies.
A second important result of this paper concerns the eccentricity evolution of binaries. We perform the first systematic study of the dependence of the evolution parameters on the integration-time cutoff and show that the reported circularization of small-mass-ratio binaries in previous works is a numerical artifact caused by discarding long-lived encounters. When those encounters are retained, $K$ remains positive across the full range of mass ratios and eccentricities we explore: all binaries eccentrify under gravitational slingshots.
Our results reshape the standard picture of hard-binary evolution in astrophysical environments and point to several important directions for future work. The CoM motion and apsidal evolution we have identified have nontrivial consequences for loss-cone refilling, the final parsec problem and the demographics of GW sources. More broadly, these results apply beyond SMBH binaries---for instance, to IMRIs, EMRIs, or stellar-mass binaries embedded in gaseous or other media. A natural next step is to incorporate these secular effects into realistic astrophysical scenarios through dedicated studies.
%%%%%%%%%%%%%%%%%%%%%%%%%%%%
\section*{Acknowledgements}
%%%%%%%%%%%%%%%%%%%%%%%%%%%%
We thank Roman Rafikov and Scott Tremaine for helpful discussions. G.M.T.\ gratefully acknowledges support from the Rubicon Fellowship, awarded by the Netherlands Organisation for Scientific Research (NWO), Grant ID \href{https://doi.org/10.61686/WYKDB06497}{https://doi.org/10.61686/WYKDB06497}. T.S. would like to thank the Institute for Advanced Study for its hospitality during the final stages of this work. T.S.\ acknowledges support from a Royal Society University Research Fellowship (URF-R1-231065). Most numerical computations used in this work were run on the IAS Typhon cluster.
\appendix

%%%%%%%%%%%%%%%%%%%%%%%%%%%%%%%%%%%%%%
\section{Numerical convergence tests}
\label{app:convergence}
%%%%%%%%%%%%%%%%%%%%%%%%%%%%%%%%%%%%%%
As detailed in Sec.~\ref{sec:num-int}, we obtain the results in this work with a code based on an explicit Runge--Kutta (\texttt{RK45}) method with an adaptive time step. A validation of our code is indirectly provided by comparisons with previous literature (see, e.g., Figs.~\ref{fig:H-K} and~\ref{fig:bonetti-comparison}). Besides such comparisons, we performed a number of numerical tests to ensure the stability and convergence of our code. We briefly report on them here.
%%%%%%%%%%%%%%%%%%%%%%%%%%%%%%%%%%%%%%%%%%
\subsection{Convergence of \texttt{RK45}}
%%%%%%%%%%%%%%%%%%%%%%%%%%%%%%%%%%%%%%%%%%

\begin{figure*}
\centering
\includegraphics[width=\textwidth]{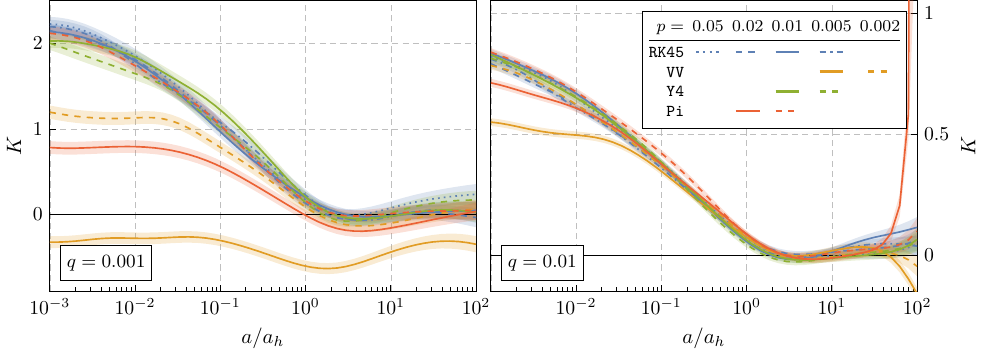}
\caption{Eccentricity growth rate $K$ as a function of $a/a_h$ for $e=0.6$ and $q=0.001$ (\emph{left panel}) or $q=0.01$ (\emph{right panel}), computed with different numerical integrators and adaptive-step thresholds $p$, as reported in the legend.}
\label{fig:convergence}
\end{figure*}

The time step in our \texttt{RK45} is chosen such that the relative change of the particle's velocity between consecutive time steps, $\abs{\Delta\vec v}/\abs{\vec v}$, does not exceed a given threshold $p$. As a default setting, we use $p=0.01$. We tested convergence by varying $p$ in the range $0.002\leq p\leq 0.05$. We focus on the eccentricity growth rate $K$, which we find to be the evolution parameter most sensitive to numerical accuracy, and consider small mass ratios for which convergence is harder to achieve. Figure~\ref{fig:convergence} shows the resulting \texttt{RK45} curves together with their $1\sigma$ statistical error bands: across the tested range of $p$, the pointwise spread between \texttt{RK45} runs is comparable to this uncertainty, with a typical $|\Delta K|/\sigma \lesssim 1$ at each $a/a_h$. A purely statistical difference between two runs would itself scatter at $\sim\sqrt{2}\sigma$, and we find no systematic trend with $p$. We therefore consider the \texttt{RK45} results shown in Fig.~\ref{fig:convergence}---and in particular those obtained at our default $p=0.01$, used throughout this work---to be converged within the uncertainty of our scattering experiments.
%%%%%%%%%%%%%%%%%%%%%%%%%%%%%%%%
\subsection{Other integrators}
%%%%%%%%%%%%%%%%%%%%%%%%%%%%%%%%
We also tested other numerical routines, including:
\begin{itemize}
\item Velocity Verlet (\texttt{VV}) \cite{PhysRev.159.98} and Yoshida 4th order (\texttt{Y4}) \cite{Yoshida:1990zz};
\item Pihajoki's integrator (\texttt{Pi}) \cite{2015CeMDA.121..211P};
\item Bulirsch-Stoer (\texttt{BS}) \cite{BulirschStoer:1966}, the same integrator used in \cite{Bonetti:2020iwk}.
\end{itemize}
\texttt{VV}, \texttt{Y4}, and \texttt{Pi} are all symplectic integrators. However, this property is broken in \texttt{VV} and \texttt{Y4} when an adaptive time step is used, which is necessary in our case. On the other hand, \texttt{Pi} is designed to support this possibility without breaking its symplectic structure.
We report the results for $K$ calculated using these integrators, for different numerical thresholds $p$, in Fig.~\ref{fig:convergence} as well. We omit \texttt{BS} from Fig.~\ref{fig:convergence}, as we found it produced much worse results than the other integrators.
%%%%%%%%%%%%%%%%%%%%%%%%%%%%%%%%%%%%%%%%%%%%%
\section{Effective minimum impact parameter}
\label{app:bmin}
\begin{figure}
\centering
\includegraphics[width=0.48\textwidth]{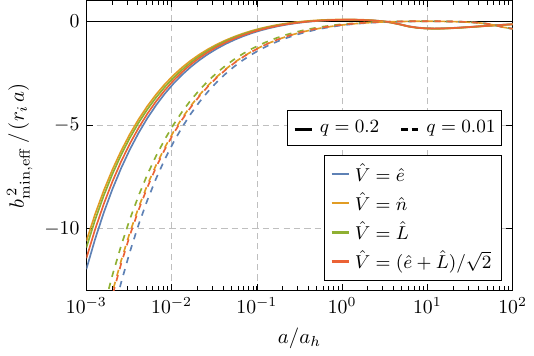}
\caption{Effective squared minimum impact parameter $b\ped{min,eff}^2$ for particles contributing to dynamical friction on a binary with CoM velocity $V\ll\sigma$. We use $e=0.6$, two values of the mass ratio ($q=0.2$ and $q=0.01$), and different CoM velocity directions. Results are obtained by matching Eq.~\eqref{eqn:P-small-V} (linear fit to the acceleration parameter from three-body experiments) with Eq.~\eqref{eqn:gamma}, and reading $b\ped{min,eff}^2$ from the linearization \eqref{eqn:bmin-sq-linear} of the integral \eqref{eqn:logL-bmin} with $b\ped{max}=r_i$. Negative $b\ped{min,eff}^2$ indicates that the inferred $\braket{\log\Lambda}$ exceeds the point-particle value $\braket{\log\Lambda}\ped{max}\approx 0.673$.}
\label{fig:bmin}
\end{figure}

%%%%%%%%%%%%%%%%%%%%%%%%%%%%%%%%%%%%%%%%%%%%%
In Sec.~\ref{sec:drag} we show how the effective Coulomb logarithm $\braket{\log\Lambda}$ can be extracted from the acceleration parameter $\vec P(\vec V)$. Here, we re-express $\braket{\log\Lambda}$ in terms of an effective minimum impact parameter $b\ped{min,eff}$, defined by substituting $\log\Lambda(v)$ with its point-mass expression,
\beq
\braket{\log\Lambda}\equiv\int_0^{\infty}\frac{e^{-v^2/2\sigma^2}}2\log\bigg[\frac{b\ped{max}^2+b_{90}^2(v)}{b\ped{min,eff}^2+b_{90}^2(v)}\bigg]\,\frac{v\dd v}{\sigma^2}\,.
\label{eqn:logL-bmin}
\eeq
The parameter $b\ped{max}$ is fixed by matching it to the value assumed in the numerical determination of $\vec P(\vec V)$. Therefore, we can solve \eqref{eqn:logL-bmin} to find $b\ped{min,eff}$.
For real values of $b\ped{min,eff}$, the maximum effective Coulomb logarithm is $\braket{\log\Lambda}\ped{max}\approx0.673$, obtained for $b\ped{min,eff}=0$. In our three-body scattering experiments, however, we measure values of $\braket{\log\Lambda}$ slightly larger than this maximum. This is not a problem per se---it means that the drag on a binary can be larger than that on a point particle. Extracting the value of $b\ped{min,eff}^2$, however, becomes more subtle, because for any $b\ped{min,eff}^2<0$, the argument of the logarithm in \eqref{eqn:logL-bmin} becomes negative for large enough $v$.
We resolve this by linearizing \eqref{eqn:logL-bmin} around $b\ped{min,eff}^2=0$, obtaining
\beq
\frac{b\ped{min,eff}^2}{r_ia}=\frac{(1+q)^2}{q}\,\frac{a_h}{a}\,\bigl[\langle\log\Lambda\rangle_{\max}-\langle\log\Lambda\rangle\bigr]\,,
\label{eqn:bmin-sq-linear}
\eeq
which is well-defined for any sign of $b\ped{min,eff}^2$. We show in Fig.~\ref{fig:bmin} the value of $b\ped{min,eff}^2$ obtained by solving \eqref{eqn:bmin-sq-linear}. As expected, $b\ped{min,eff}^2$ is of order $\mathcal O(r_ia)$, implying that the correction from the point-mass result comes from particles whose pericenter distance is of order $\mathcal O(a)$ from the binary's CoM. The value of $b\ped{min,eff}^2$ is negative across most values of $a/a_h$, and grows in magnitude for very hard binaries.
We emphasize that, in any case, such nonzero value of $b\ped{min,eff}^2$ corresponds to only a small correction to $\braket{\log\Lambda}$. This is easily seen in \eqref{eqn:bmin-sq-linear}, where the hardness factor $a/a_h\ll1$ enters the relation between the two quantities.
%

%%%%%%%%%%%%%%%%%%%%%%%%%%%%%%%%%%%%%%%%%%%%%%%%%%%
\section{Eccentricity growth at small mass ratios}
\label{app:comparison}
\begin{figure}
\centering
\includegraphics[width=0.49\textwidth]{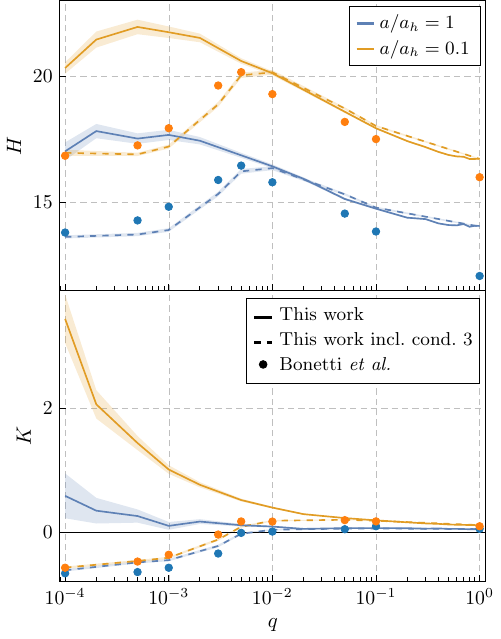}
\caption{Hardening rate $H$ (\emph{top panel}) and eccentricity growth rate $K$ (\emph{bottom panel}) as functions of mass ratio $q$ for $e=0.6$. Solid and dashed curves show results from our code using the Hubble-time integration cutoff~\eqref{eqn:Nmax-Hubble}: dashed lines include the additional stopping condition 3 of Refs.~\cite{Rasskazov_2019,Bonetti:2020iwk}, while solid lines do not. The colored markers show data digitized from Fig.~1 of Ref.~\cite{Bonetti:2020iwk}. When condition 3 is imposed, our results reproduce the negative $K$ at small $q$ reported in those works.}
\label{fig:bonetti-comparison}
\end{figure}

%%%%%%%%%%%%%%%%%%%%%%%%%%%%%%%%%%%%%%%%%%%%%%%%%%%
In this appendix we identify the specific methodological difference responsible for the negative values of the eccentricity growth rate $K$ at small mass ratios reported by Rasskazov \emph{et al.}~\cite{Rasskazov_2019} and Bonetti \emph{et al.}~\cite{Bonetti:2020iwk}, and demonstrate that removing it yields $K>0$.
Both Refs.~\cite{Rasskazov_2019,Bonetti:2020iwk} employ three stopping criteria for their three-body scattering experiments:
\begin{enumerate}
\item the star exits the sphere of radius $r\ped{sph}=50a$ with positive energy;
\item the elapsed integration time exceeds a maximum value $T\ped{max}$;
\item the cumulative time spent by the particle inside the sphere of radius $r\ped{sph}=50a$ exceeds $\num{2e4}$ binary orbital periods.
\end{enumerate}
Only scatterings satisfying condition 1 are considered resolved and contribute to the measured rates. Particles removed by conditions 2 or 3 are instead discarded.
The first difference between our setup and those works concerns the maximum integration time $T\ped{max}$. We adopt a universal cutoff $N\ped{max} = 10^{11}/(2\pi)$ (see Sec.~\ref{sec:num-int}), sufficiently large such that only a small fraction of scatterings remain unresolved (Tab.~\ref{tab:resolved-fractions}). Refs.~\cite{Rasskazov_2019,Bonetti:2020iwk} instead adopt a physical cutoff of a Hubble time~\eqref{eqn:Nmax-Hubble}.\footnote{Note that the fiducial parameters in Ref.~\cite{Rasskazov_2019} ($M=4\times10^6 M_\odot$, $\sigma=70 \mathrm{km/s}$) are not fully consistent with the $M$--$\sigma$ relation~\cite{Kormendy:2013dxa}, which predicts $\sigma\approx99 \mathrm{km/s}$ for that mass. This reduces the effective cutoff in~\cite{Rasskazov_2019} by roughly a factor of $3$.} For most mass ratios ($q \gtrsim 5\times 10^{-4}$), our cutoff time exceeds theirs. However, $K$ already becomes negative at $q \sim 5\times 10^{-3}$ in~\cite{Rasskazov_2019,Bonetti:2020iwk}, a regime where our cutoff is still larger and where $K$ is well converged in our simulations (see Fig.~\ref{fig:HKPQ-Tmax}). Differences in $T\ped{max}$ alone therefore cannot explain the sign change.
The key difference is condition 3, which removes particles that spend longer than $\num{2e4}$ binary orbital periods inside the sphere $r<r\ped{sph}$.\footnote{We note that condition 3 was likely motivated by computational cost. References~\cite{Rasskazov_2019,Bonetti:2020iwk} integrate the full three-body problem, in which particles on long-lived metastable orbits may reach very large distances, introducing disparate scales that are expensive to resolve. In the restricted three-body problem adopted in this work, long-lived encounters are no more expensive per unit time than short ones.} To isolate its effect, we perform a controlled comparison. Starting from our code with $T\ped{max}$ set to the Hubble-time value of Eq.~\eqref{eqn:Nmax-Hubble} (identical to~\cite{Rasskazov_2019,Bonetti:2020iwk}), we run two versions of the code: one imposing condition 3 (as in~\cite{Rasskazov_2019,Bonetti:2020iwk}), and one without it.\footnote{As reported throughout, we typically set $N\ped{max} = 10^{11}/(2\pi)$ in our simulations. However, to truly isolate the effect of condition 3, this is the one instance in which we extend $N\ped{max}$ beyond $10^{11}/(2\pi)$, since for $q \lesssim 5 \times 10^{-4}$ the Hubble-time cutoff $N_H$ exceeds our default $N\ped{max}$.}
Figure~\ref{fig:bonetti-comparison} shows $H$ and $K$ as functions of $q$ at fixed $a/a_h = 1$ and $a/a_h = 0.1$, for $e = 0.6$. The colored markers denote data points digitized from Fig.~1 of Bonetti \emph{et al.}~\cite{Bonetti:2020iwk}; the dashed lines show our code with condition 3 imposed; solid lines show our code without condition 3. When condition 3 is imposed, our results reproduce those from~\cite{Rasskazov_2019,Bonetti:2020iwk} very well, including the negative $K$ at small $q$. When it is removed, $K$ remains positive across the full mass-ratio range, confirming that condition 3 is the source of the negative $K$ values. 

\begin{table}
\centering
\begin{tabular}{crrrrrrrr}
\toprule
(\%) & \multicolumn{2}{c}{$N\ped{max}=\frac{10^{5}}{2\pi}$}& \multicolumn{2}{c}{$N\ped{max}=\frac{10^{8}}{2\pi}$}& \multicolumn{2}{c}{$N\ped{max}=\frac{10^{11}}{2\pi}$}& \multicolumn{2}{c}{Incl.~cond.~3} \\
\cmidrule(lr){2-3}\cmidrule(lr){4-5}\cmidrule(lr){6-7}\cmidrule(lr){8-9}
$q$ & mean & worst & mean & worst & mean & worst & mean & worst \\
\midrule
0.0001 & 77.8 & 48.6 & 85.2 & 61.3 & 96.0 & 84.3 & 94.9 & 82.1 \\
0.001 & 78.6 & 50.2 & 91.6 & 74.1 & 98.7 & 93.3 & 97.3 & 90.1 \\
0.01 & 85.3 & 61.2 & 96.0 & 84.1 & 99.8 & 98.4 & 98.8 & 96.0 \\
0.1 & 91.3 & 72.8 & 98.5 & 92.8 & $\num{>99.9}$ & 99.8 & 99.4 & 97.9 \\
1 & 93.8 & 78.8 & 99.3 & 96.3 & $\num{>99.9}$ & 99.9 & 99.7 & 98.5 \\
\bottomrule
\end{tabular}
\caption{Resolved fraction of scattering experiments for $e = 0.6$. ``Mean'' denotes the average over velocity bins; ``worst'' is the minimum over bins (typically the smallest velocity bin). The first three column pairs show results for increasing values of $N\ped{max}$. The last column pair shows $N\ped{max} = 10^{11}/(2\pi)$ with the additional stopping condition 3 of Refs.~\cite{Rasskazov_2019,Bonetti:2020iwk}.}
\label{tab:resolved-fractions}
\end{table}

It may seem surprising that discarding only a small fraction of encounters can have such a large effect. Indeed, Tab.~\ref{tab:resolved-fractions} shows that only a few percent of encounters are removed by condition 3. However, $K$ arises from a near-cancellation between positive and negative contributions to the angular-momentum exchange, and is therefore especially sensitive to the removal of any subpopulation with systematically different properties. This is also reflected in the statistical uncertainties: at small mass ratios, the solid curves have visibly larger error bars than the dashed ones. This indicates that the encounters removed by condition 3 carry comparatively large individual changes in energy and angular momentum and contribute disproportionately to the variance of the sample averages. Bonetti \emph{et al.}~\cite{Bonetti:2020iwk} interpreted the negative $K$ at small $q$ in terms of a decomposition into quick/late and prograde/retrograde encounters, concluding that late-retrograde encounters circularize the binary. While that decomposition may still correctly identify which encounters are most affected by the cutoff, Fig.~\ref{fig:bonetti-comparison} shows that the physical conclusion drawn from it is misleading: once condition 3 is removed, $K$ remains positive for all mass ratios.
%

%%%%%%%%%%%%%%%%%%%%%%%%%%%%%%%%%%%%%%%%%%%%%%%%
\section{Uncertainties on the binary evolution}
\label{app:uncertainties}
%%%%%%%%%%%%%%%%%%%%%%%%%%%%%%%%%%%%%%%%%%%%%%%%
In Sec.~\ref{sec:eoms} we solve the equations of motion for the binary state vector $\vec\psi=(t, e, V_x, V_y, \varpi, x, y)^\intercal$. These are given by equations \eqref{eqn:dtdxi}, \eqref{eqn:dedxi}, \eqref{eqn:dVdxi}, \eqref{eqn:dvarpidxi}, supplemented with $\dd\vec r/\dd t=\vec V$, and can be written schematically as
\beq
\frac{\dd\vec\psi}{\dd\xi}=\vec\Upsilon(\vec\psi,\xi)\,,
\label{eqn:dpsidxi}
\eeq
where $\xi\equiv\log(a_h/a)$ and $\vec\Upsilon$ depends on the rates measured through three-body scattering experiments, namely $H$, $K$, $\vec P$, and $Q$. We compute the right-hand side $\vec\Upsilon$ on a discrete grid of eccentricity values, labeled by $e_i$ with $i=1,\ldots,N_e$, each of which requires an independent set of three-body scattering simulations. We then interpolate the data using a four-point Lagrange stencil. More precisely, for a given value of the eccentricity $e$ at runtime, we evaluate $\vec\Upsilon$ by assigning a weight $w_i(e)$ to the $i$-th set of data (all of which are zero, except for those corresponding to the four grid points closest to $e$), and taking the weighted average.
The right-hand side of \eqref{eqn:dpsidxi} carries a numerical uncertainty $\delta\vec\Upsilon$. Besides integrating Eq.~\eqref{eqn:dpsidxi} to find the length-7 vector $\vec\psi(\xi)$, we therefore wish to determine the $7\times7$ covariance matrix $\mathsf{C}(\xi)$ of $\vec\psi(\xi)$ induced by $\delta\vec\Upsilon$. We proceed as follows. 
Varying Eq.~\eqref{eqn:dpsidxi} gives
\beq
\frac{\dd(\delta\vec\psi)}{\dd\xi}=\mathsf{J}(\vec\psi,\xi)\delta\vec\psi+\delta\vec\Upsilon\,,\qquad\mathsf{J}\equiv\frac{\partial\vec\Upsilon}{\partial\vec\psi}\,.
\label{eqn:variation}
\eeq
The first term describes the feedback of perturbations of $\vec\psi$ through the Jacobian $\mathsf{J}$, while the second term is the direct forcing from the uncertainties on the rates. Data from different eccentricity grid points arise from independent sets of scattering simulations and are therefore uncorrelated. Furthermore, at a given eccentricity grid point, we approximate the five rates $H$, $K$, $P_{\hat e}$, $P_{\hat n}$, and $Q$ as having uncorrelated uncertainties. This is a minor approximation that allows us to reduce the storage needed for the three-body scattering data. 
We then introduce independent standard normals $\eta_{ij}$ with $i=1,\ldots,N_e$ and $j=1,\ldots,5$, that is, one for each rate at each eccentricity grid point available in the data. The variation of the state vector can then be written as a linear combination of $\eta_{ij}$ with weights $\vec F_{ij}$,
\beq
\delta\vec\psi=\sum_{ij}\vec F_{ij}\eta_{ij}\,.
\label{eqn:dpsi}
\eeq
Given the uncertainty $\sigma_{ij}$ on the rate $j$ at the $i$-th eccentricity grid point, the uncertainty on $\vec\Upsilon$ follows by error propagation,
\beq
\delta\vec\Upsilon=\sum_{ij}w_i(e)\vec B_j\sigma_{ij}\eta_{ij}\,,
\label{eqn:dUpsilon}
\eeq
where $\vec B_j=\partial\vec\Upsilon/\partial\mathcal R_j$ encodes the dependence of $\vec\Upsilon$ on the rate vector $\mathcal R=\{H,K,P_{\hat e},P_{\hat n},Q\}^\intercal$.
Substituting Eqs.~\eqref{eqn:dpsi} and \eqref{eqn:dUpsilon} into Eq.~\eqref{eqn:variation} yields
\beq
\frac{\dd\vec F_{ij}}{\dd\xi}=\mathsf J\,\vec F_{ij}+w_i(e)\vec B_j\sigma_{ij}\,.
\eeq
We thus have $5\times N_e$ coupled differential equations, each for a 7-dimensional vector $\vec F_{ij}$, which we solve alongside \eqref{eqn:dpsidxi} with the initial condition $\vec F_{ij}(0)=\vec 0$. The weights $w_i(e)$ ensure that grid points outside the active stencil contribute no instantaneous forcing, though their $\vec F_{ij}$ still evolve due to the uncertainty on the state vector $\vec\psi$.
The covariance matrix of $\vec\psi$ follows from Eq.~\eqref{eqn:dpsi} as
\beq
\mathsf{C}=\mathbb{E}[\delta\vec\psi\,\delta\vec\psi^\intercal]=\sum_{ij}\vec F_{ij}\vec F_{ij}^\intercal\,.
\eeq
The $1\sigma$ bands in Fig.~\ref{fig:evolution} are obtained from the diagonal elements of $\mathsf{C}$, while the ellipses drawn in Fig.~\ref{fig:trajectory} use the $2\times2$ covariance matrix between $x$ and $y$.
The entries of the Jacobian $\mathsf{J}$ require computing partial derivatives of $\vec\Upsilon$. We obtain the derivatives with respect to $e$ analytically from the Lagrange weights $w_i(e)$ and the interpolated rates. The remaining derivatives---with respect to $V_x$, $V_y$, and $\varpi$---are computed numerically by finite differences.
%%%%%%%%%%%%%%%%%%%%%%%%%%%%%%%
\section{$N$-body simulations}
\label{app:Nbody}
%%%%%%%%%%%%%%%%%%%%%%%%%%%%%%%
\begin{figure*}
\centering
\includegraphics[width=\textwidth]{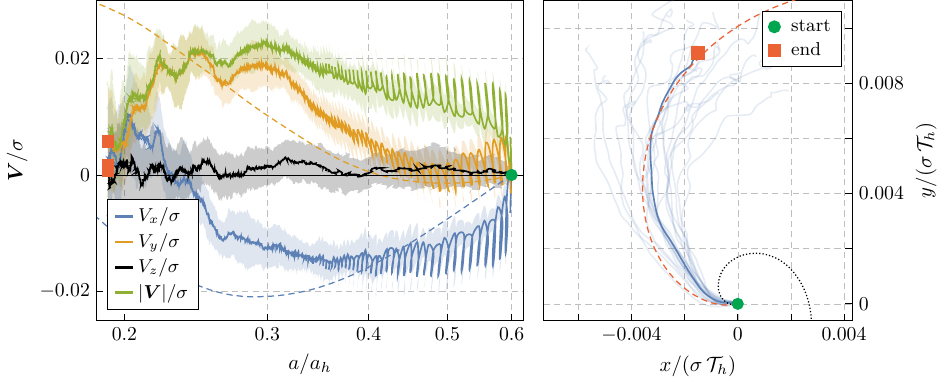}
\caption{Results of the $N$-body simulations for a binary with the same parameters as in Figs.~\ref{fig:evolution} and \ref{fig:trajectory} ($q=0.2$, $e_0=0.5$, and $a_0/a_h=0.6$). \emph{Left panel}: components of the CoM velocity. The solid curves show the $N$-body ensemble mean over 20 realizations, and the shaded band denotes the $1\sigma$ scatter across realizations. The dotted lines indicate $V_x$ and $V_y$ as obtained from the three-body simulations, with a modified outer cutoff $b\ped{max,eff}$ in the dynamical-friction term~\eqref{eqn:Pdf}, chosen to account for the finite size of the $N$-body box (see text). \emph{Right panel}: trajectory of the CoM in the orbital plane. Individual realizations are shown as low-opacity curves, while the thick line denotes the ensemble mean. The red dashed line shows the three-body prediction with the $b\ped{max,eff}$ cutoff, while the black dotted line uses the unmodified dynamical-friction expression and coincides with the trajectory shown in Fig.~\ref{fig:trajectory}.}
\label{fig:nbody-validation}
\end{figure*}
To validate the results presented in Sec.~\ref{sec:binary_evolution}, we perform direct $N$-body simulations of a binary immersed in a uniform background of lighter particles.
We consider a Keplerian binary with total mass $M$, mass ratio $q$, initial semi-major axis $a_0$, and eccentricity $e_0$, placed at the center of a cubic box of side length $L$ with periodic boundary conditions. The binary is initialized at periapsis, with its eccentricity vector aligned with the $\hat x $ direction and its angular momentum aligned with $\hat z$, so that the orbit lies in the $xy$ plane. The background consists of $N$ equal-mass particles, each with mass $m = \rho L^3 / N$ distributed uniformly in space. Their initial velocities are drawn from a Maxwellian distribution~\eqref{eqn:maxwellian-1d}.
To reduce computational cost, we adopt a \emph{restricted} $N$-body approach in which background particles do not interact with each other. Each particle moves only in the gravitational field of the two binary components, while each binary component feels the gravity of its companion and of all particles. This reduces the computational cost to $\mathcal{O}(N)$ per time step, allowing us to simulate large particle numbers, $N \sim \mathcal{O}(10^6)$, while still capturing the processes responsible for binary hardening, the CoM force, and precession. The system is evolved using a second-order symplectic leapfrog integrator with fixed time step $\Delta t$.
We soften the gravitational interaction between each background particle and binary component as in Eq.~\eqref{eq:softening}, with softening length $\epsilon = 5\times 10^{-3}\,a_0$. Periodic boundary conditions are implemented via the minimum-image convention, applied to stellar positions relative to the binary’s instantaneous CoM. While the background is initially uniform, stars ejected by the binary re-enter through the periodic boundaries, so that the mean density remains constant throughout the simulation.
Because the background is composed of a finite number of particles, the binary experiences not only the secular, deterministic evolution driven by the average outcome of three-body interactions, but also stochastic fluctuations (Brownian motion). To disentangle these contributions, we run an ensemble of $N\ped{sim}$ simulations with identical physical parameters but different random realizations of the background. We then compute the mean evolution and its variance across runs. The stochastic component averages down as $1/\sqrt{N\ped{sim}}$, while the secular component persists.
We choose $q = 0.2$, $e_0 = 0.5$, and $a_0/a_h = 0.6$, matching the parameters used for the evolution shown in Figs.~\ref{fig:evolution} and~\ref{fig:trajectory}. The background is composed of $N = 5 \times 10^6$ particles in a cubic box of side length $L = 38\,a_0$, chosen such that $L \gg a_h$. This corresponds to a particle mass $m = 5.6 \times 10^{-7} M$. The integration time step is $\Delta t = 5 \times 10^{-3}\,T_0$, where $T_0$ is the initial orbital period. We evolve $N\ped{sim}=20$ independent realizations of the system.
Figure~\ref{fig:nbody-validation} summarizes the results. The left panel shows the components of the CoM velocity. The out-of-plane component $V_z$ remains consistent with zero throughout the simulation, in agreement with the symmetry argument of Sec.~\ref{sec:symmetry}. The right panel shows the trajectory of the binary's CoM in the orbital plane, where the binary follows the same outward spiral described in Sec.~\ref{sec:binary_evolution}.
For comparison, we overlay the predictions from the three-body calculations (dashed lines), obtained by solving the evolution equations~\eqref{eqn:dedxi}-\eqref{eqn:dvarpidxi}. Although the two trajectories agree qualitatively, the characteristic radius obtained in the $N$-body simulation is larger than that obtained from three-body scattering. This might be due to the finite size of the simulation box, which limits the size of the wake responsible for dynamical friction. To mimic this effect, we reduce the outer cutoff entering \eqref{eqn:logLambda} from its nominal value, $b\ped{max}=r_i=4a_h(1+q)^2/q$, to an effective scale $b\ped{max,eff}=\sqrt{a r_i}$, which we use in Fig.~\ref{fig:nbody-validation}. With this single modification, we find good agreement between the three-body and $N$-body calculations, both in the overall evolution of the CoM velocity and in the outward-spiraling trajectory. In the right panel, we additionally show the three-body solution with the usual cutoff on dynamical friction (black dotted line). This corresponds to the same trajectory shown in Fig.~\ref{fig:trajectory}.
Remarkably, this minimal $N$-body setup reproduces the key conclusions of the three-body framework: the CoM develops a nonzero in-plane velocity and follows an outward-spiraling trajectory.

\clearpage
\bibliography{main}% Produces the bibliography via BibTeX.

\end{document}